\pdfoutput=1

\documentclass[3p,times,twocolumn,sort&compress]{elsarticle}

\usepackage{ecrc-ac}
\usepackage[british]{babel}
\usepackage{ulem}

\usepackage{scalefnt}

\volume{00}

\firstpage{1}

\journalname{Nuclear Physics B Proceedings Supplement}

\runauth{K.G.~Chetyrkin, J.H.~K\"uhn, M.~Steinhauser, C.~Sturm}

\jnltitlelogo{sfb-logo}

\jid{nuphbp}

\jnltitlelogo{Nuclear Physics B Proceedings Supplement}

\usepackage{amssymb}
\usepackage{framed,color}
\usepackage{stackrel}
\usepackage{hyperref}
\usepackage{graphicx}
\usepackage{wasysym}
\usepackage{amssymb}
\graphicspath{{./}{./figures/}}

\usepackage[figuresright]{rotating}

\newcommand{\ice}[1]{\relax}

\newcommand{\MSbar}{\overline{\mbox{MS}}}

\newcommand{\ri}{\mbox{i}}
\newcommand{\varep}{\varepsilon}
\newcommand{\D}{\mbox{d}}

\begin{document}


\begin{frontmatter}



\dochead{}

\title{Massive Tadpoles: Techniques and Applications}


\author[KA]{Konstantin G. Chetyrkin}
\author[KA]{Johann H. K\"uhn}
\author[KA]{Matthias Steinhauser\corref{cor1}}\ead{matthias.steinhauser@kit.edu}
\address[KA]{Institut f\"ur Theoretische Teilchenphysik, Karlsruher
  Institut f\"ur  Technologie (KIT), Wolfgang-Gaede-Stra{\ss}e 1, D-76128 Karlsruhe, Germany}

\cortext[cor1]{Corresponding author}

\author[Wue]{Christian Sturm}
\address[Wue]{Universit{\"a}t W{\"u}rzburg, Institut f{\"u}r Theoretische Physik und Astrophysik, Emil-Hilb-Weg 22, D-97074 W{\"u}rzburg, Germany}

\begin{abstract}
  A brief discussion of massive tadpole diagrams and their phenomenological
  consequences is presented. This includes predictions of the $\rho$ parameter
  and, as a consequence, the mass of the $W$ boson, implications on the charm
  and bottom quark masses from the moments, i.e. the derivatives of the
  current correlators, and the Higgs boson decay rate. A fairly consistent
  picture emerges, with $m_c(3~\mbox{GeV})=0.986\pm0.013$~GeV and 
  $m_b(m_b)=3.610\pm0.016$~GeV. Furthermore, fairly stringent predictions for the
  Higgs decay rate into photons and gluons are obtained, which will be
  interesting in increasingly precise experiments.
\end{abstract}


\end{frontmatter}




\section{Introduction\label{sec:Introduction}}

Using massive tadpole diagrams significant progress has been made in the
improved prediction of various physical quantities. They have, e.g.,
contributed to an amazingly precise
relation between quark masses (in particular the
top-quark mass $M_t$), the mass of the Higgs boson, $M_H$, and the
masses of the gauge bosons, $M_W$ and $M_Z$. 
Three- and even four-loop
corrections have become accessible during the past years. Direct and
indirect measurements are well consistent, at least within the current
world average $M_t=173.21\pm0.51\pm0.71$~GeV~\cite{Agashe:2014kda} and
$M_H=125.7\pm0.4$~GeV~\cite{Agashe:2014kda}.  Current-current
correlators, evaluated in three- and partially four-loop approximation
are thus an indispensable tool for tests of the Standard Model (SM),
as we will discuss in more detail in Section~\ref{sec:II}.

The evaluation of three- and even four-loop tadpole diagrams is
directly related to the evaluation of moments of the charm- and
bottom-quark correlators. These may in turn lead to a precise
determination of charm- and bottom-quark masses.  Since all these
quantities, in turn, are directly accessible, both to a perturbative
and a non-perturbative treatment, a remarkably consistent picture
emerges. In fact the analysis for the charm- as well as the
bottom-quark mass leads to $m_c(3~\mbox{GeV})=0.986\pm{0.013}$~{GeV} and
$m_b(10~\mbox{GeV})=3.610\pm{0.016}$~GeV~\cite{Kuhn:2007vp,Chetyrkin:2009fv},
a result quite comparable to other methods, in particular to 
non-perturbative studies as will be 
discussed in Section~\ref{sec:III}.

Finally, in Section~\ref{sec:IV}, we list a collection of topics and
results which are connected to the main theme of this
article. This includes the decoupling of heavy
quarks at four loops, the Higgs-gluon coupling up to five-loop order,
and the Higgs-decay rate into two photons 
at three and four loops, including non-singlet and singlet terms.


\section{Weak corrections\label{sec:II}}

Let us start with the $\rho$ parameter, the quantity introduced in the
early times of electroweak interactions of the SM. In its more modern
version it gives the relation between gauge boson masses, the weak-
and the electromagnetic coupling $G_F$ and $\alpha$, a relation, which
exists at tree level already. In higher orders this relation depends
on all remaining parameters of the SM. The radiative corrections are
dominated by the quadratic dependence on the mass of the top quark
$M_t$, the logarithmic dependence on the Higgs boson, $M_H$, and,
to a lesser extent, also on the masses of the remaining quarks
$m_q$ and leptons $m_{\ell}$:
\begin{equation}
  M_W=f(G_F, M_Z,\alpha; M_t, M_H; m_q, m_{\ell} ).
\end{equation}
A slightly different version of the same equation 
\begin{equation}
M_W^2\left(1-{M_W^2\over
  M_Z^2}\right)={\pi\alpha\over\sqrt{2}G_F}(1+\Delta r)
\end{equation}
makes the presence of the electroweak corrections even more
transparent. This equation can be rewritten and simplified even further
by separating $\Delta r$ into a piece which is dominated by weak effects
and another one which is dominated by electromagnetic effects, mainly
due to the running of the electromagnetic coupling $\Delta\alpha$
(see Refs.~\cite{Steinhauser:1998rq} and~\cite{Sturm:2013uka} for three- and
four-loop corrections, respectively). Furthermore, it is convenient to
separate the leading $M_t^2$ dependence which leads to
\begin{equation}
  \Delta r = -{\cos^2{\theta_W}\over \sin^2{\theta_W}}\Delta\rho +
  \Delta\alpha+\Delta r_{\mbox{\tiny{remaining}}}\,.
\end{equation}
Here $\Delta\rho={3{G_F M_t^2/(8\sqrt{2}\pi^2)}}+\ldots$ incorporates the
dominant weak terms evaluated in leading order by
Veltman~\cite{Veltman:1977kh} nearly 40 years ago.

It is the aim of ongoing and of future theoretical studies  to compete with the
precision anticipated for the  next round of  experiments. The present precision
and the precision anticipated for the future  (as far
as $\delta M_W$ and $\delta M_t$ are concerned) are given by
\begin{center}
{\scalefont{0.9}
\begin{tabular}{c|c|c}
$\delta M_W$ [MeV]&$\delta M_t$ [GeV]&\\\hline
33                & 5                & status 2003 (LEP, TEV)\\\hline
15                & 0.76             & now (TEVATRON, LHC)\\\hline
8$\to$5           & 0.6              & aim (LHC); theory\\\hline
3$\to$1.2         & 0.1-0.2          & ILC, TLEP
\end{tabular}
}
\end{center}
%
As it turns out, the relative shifts of $M_W$ and $M_t$ are just of the
right order of magnitude to explore the sensitivity towards radiative
corrections. This is seen most clearly by considering the shift in $M_t$
that is compensated by a shift in $M_W$
\begin{equation}
  \delta M_W\approx 6\cdot 10^{-3}\delta M_t
  \,,
\end{equation}
keeping $\alpha(M_Z)$, $M_Z$ and $M_H$ fixed.

Let us now recall the development of the theory predictions for
$\Delta\rho$ during the past one or two decades.

\begin{figure}[!ht]
  \begin{center}
    \includegraphics[clip,width=3.5cm]{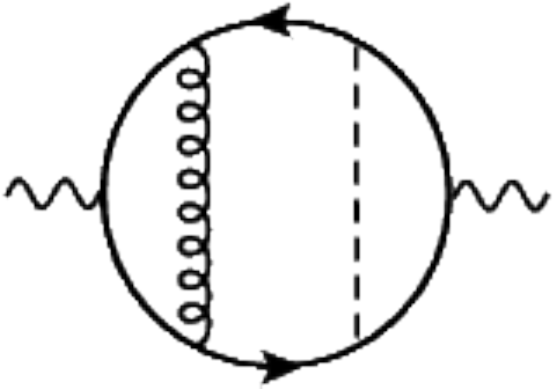}
  \end{center}
  \vspace*{-0.6cm}
  \caption{Sample diagram for the $\alpha_s X_t^2 f(M_t/M_H)$ contribution.
    Solid lines denote top quarks, dashed lines Higgs or Goldstone bosons and
    curly lines gluons.\label{fig:rhoQCD}}
\end{figure}

Early results related to the two-loop approximation can be found in
Refs.~\cite{Barbieri:1992nz,Fleischer:1993ub}.  These papers are based
on the approximation $M_t^2\gg M_W^2$.  The first step into the
three-loop regime was taken in the limit
$M_H=0$~\cite{vanderBij:2000cg}. In fact, this turns out to be a poor
approximation, leading to tiny corrections for the terms of order
$X_t^3$ and $\alpha_s X_t^2$ with
\begin{equation}
  X_t={G_F M_t^2\over8\sqrt{2}\pi^2}\,.
\end{equation}  
The first three-loop result with $M_H$
different from zero requires the full set of three-loop
tadpoles~\cite{Faisst:2003px}. At order $\alpha_s X_t^2
f(M_t/M_H)$ this corresponds to QCD corrections to the two-loop
diagrams of order $X^2_t f(M_t/M_H)$ (see Fig.~\ref{fig:rhoQCD}
  for a sample Feynman diagram). At order $X_t^3 f(M_t/M_H)$
diagrams with one quark line contribute, as well as those involving
two disconnected quark lines.

\begin{figure}[!h]
  \begin{center}
    \includegraphics[width=3.5cm]{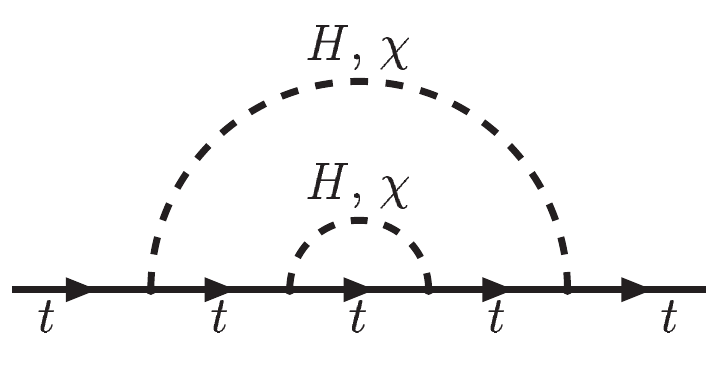}
    \includegraphics[width=3.5cm]{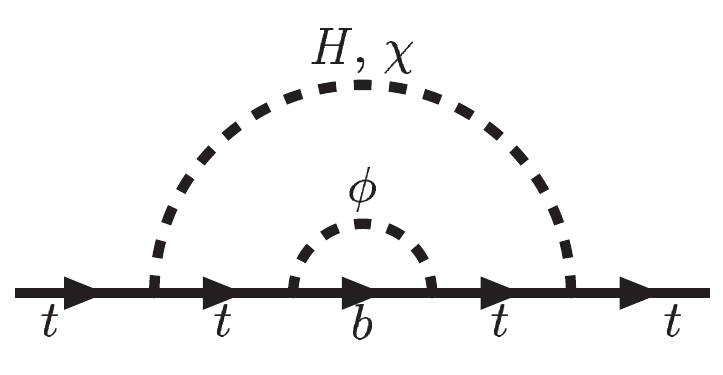}
    \includegraphics[width=3.5cm]{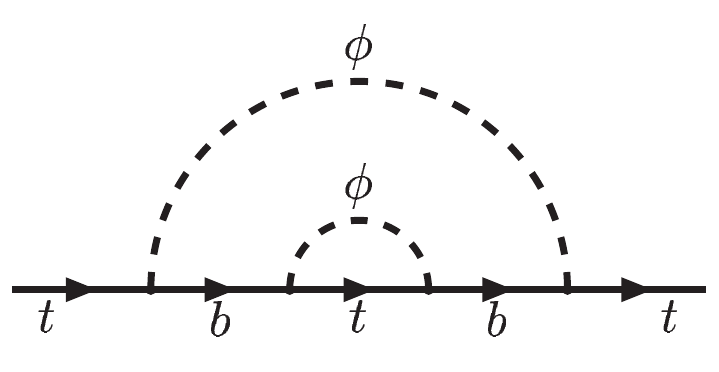}
    \includegraphics[width=3.5cm]{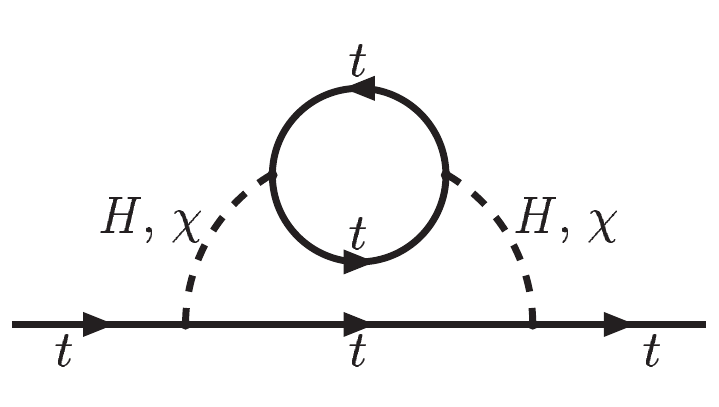}
    \includegraphics[width=3.5cm]{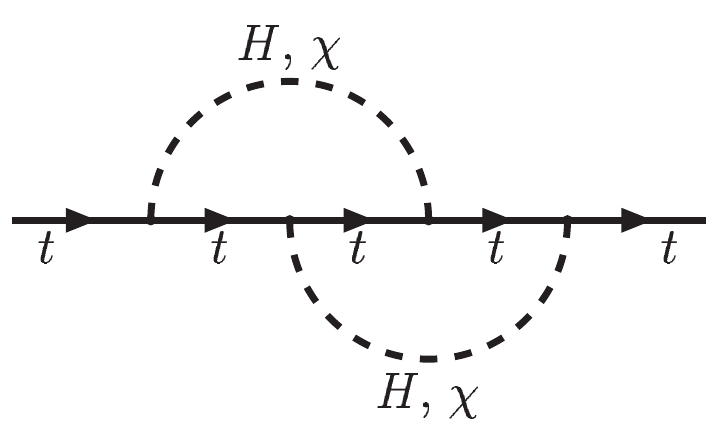}
  \end{center}
  \vspace*{-0.6cm}
  \caption{Two-loop on-shell diagrams which contribute to the translation
    of the $\MSbar$ mass to the pole mass.\label{fig:onshelldia}}
\end{figure}

At the same time the translation from the $\MSbar$ mass $m_t(M_t)$
to the pole mass $M_t$ has to be performed at two loops. This
corresponds to the evaluation of two-loop on-shell diagrams. For the case
of the $X_t^3 f(M_t/M_H)$ corrections the counterterms are of order
$X_t^2$ and are depicted in Fig.~\ref{fig:onshelldia}.

\begin{figure}[!h]
  \begin{center}
    \includegraphics[clip,width=9cm]{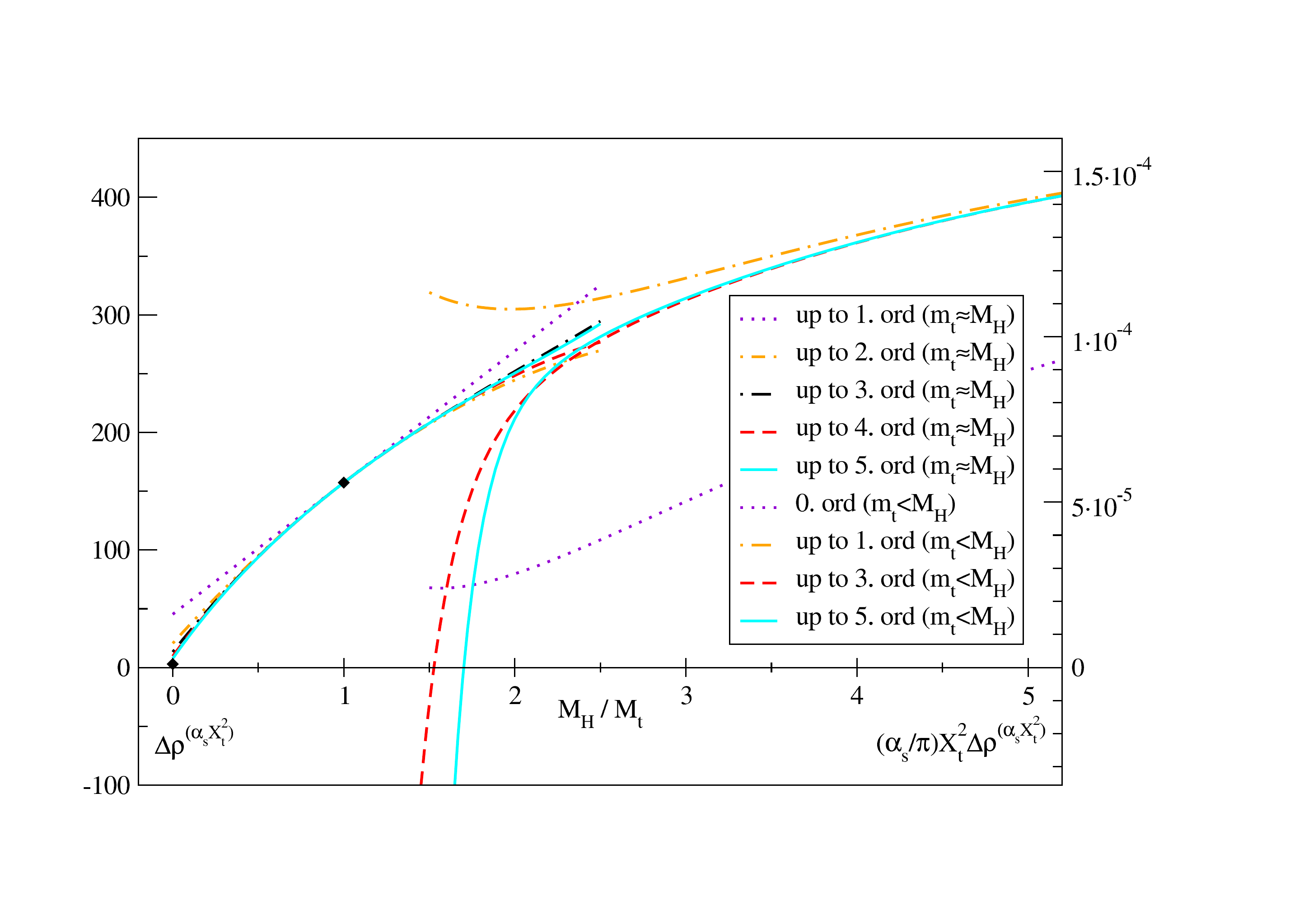}
  \end{center}
  \vspace*{-1.1cm}
  \caption{\label{fig:exprho} Contributions of order $\alpha_s X_t^2$
    to $\Delta \rho$ in the on-shell definition of the top-quark
    mass. The black squares indicate the points where the exact result
    is known. Using the latest numerical values for $M_t$ and
      $M_H$ one obtains $M_H/M_t\approx 0.73$.}
\end{figure}

In contrast to the pure QCD problem (see
Section~\ref{sec:III}) with only one mass scale being present (and
which can therefore be expressed in closed analytic form) here one
typically encounters two different scales, $M_H$ and $M_t$. A closed
analytical solution is no longer at hand. There are, however, various
cases which allow for expansions, one of which is perfectly valid for
$M_H$ and $M_t$ in the region of interest. Analytic results are
available in the cases $M_H=0$ and $M_H=M_t$, where only one scale is
present. Expansions, which in principle allow for arbitrary high
precision are then accessible in two cases: for the case of large
$M_H$ with the approximation in $(M_t^2/M_H^2)^n$ modulo logarithms
which is valid down to $M_H \approx 2 M_t$ and the case of $M_H$
around $M_t$ which is valid from fairly small $M_H$, say
$M_H\approx 0.1 M_t$, up to $M_H\approx 2M_t$.  The results for
the expansion in $(M_H/M_t)^n$ and in
$(M_H-M_t)^2/M_t^2$ are shown in Fig.~\ref{fig:exprho}. Note
that for $M_H=0/126$~GeV one obtains for the prefactor of $\alpha_s
X_t^2$, a part of the three-loop term of $\Delta \rho$, the values
$2.9/120$.

\begin{figure}[!h]
  \begin{center}
    \includegraphics[clip,width=8cm]{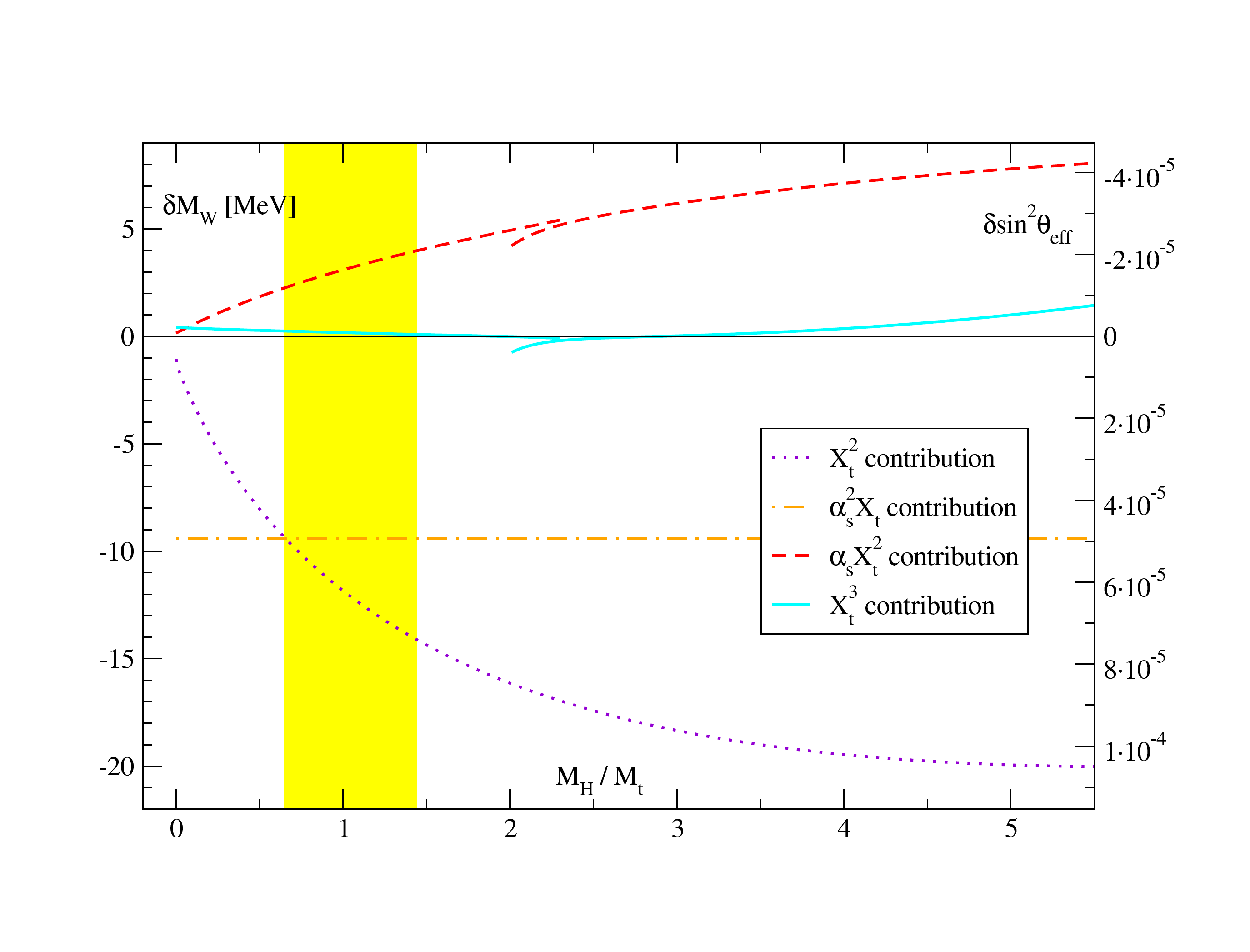}
  \end{center}
  \vspace*{-1.1cm}
  \caption{\label{fig:shift} The shift in $M_W$ and
      $\sin^2\theta_{\rm eff}$ as a function of $M_H/M_t$ induced by
      the corrections of order $X_t^2$, $\alpha_s^2 X_t$, $\alpha_s
      X_t$, and $X_t^3$.}
\end{figure}

The results for the shift in $M_W$ and the effective weak mixing
angle are shown in Fig.~\ref{fig:shift} for the
four contributions which are most relevant. The terms proportional to
$X_t$ and $X_t\alpha_s$
must be taken into account in any sensible analysis and amount to a
shift in $\delta\rho$ of order 0.00865.

The two-loop piece proportional to $X_t^2$ is of the same
order as the three loop piece, proportional to $\alpha_s^2 X_t$.  
The purely weak term proportional $X_t^3$ is negligible now and in the
foreseeable future, the term  proportional to $\alpha_s X_t^2$ is 
just below the present sensitivity.

\subsection*{Four-loop QCD contributions}

Two- and three-loop QCD corrections to $\Delta\rho$ have been
computed in
Refs.~\cite{Djouadi:1987gn,Djouadi:1987di,Kniehl:1988ie,Avdeev:1994db,Chetyrkin:1995ix,Chetyrkin:1995js}
  about 20 years ago.  As stated above, since several years it is now
possible to push the predictions for the $\rho$ parameter to the
four-loop level. This requires, on the one hand, the relation between
pole- and $\MSbar$ mass in three-loop
approximation~\cite{Chetyrkin:1999qi,Melnikov:2000qh} and, on the
other hand, the evaluation of about fifty
four-loop tadpole diagrams.  This has been achieved by a
combination of analytically methods, difference equations and
semi-numerical
integration~\cite{Chetyrkin:2006bj,Boughezal:2006xk,Faisst:2006sr}.
In a first step this has lead
to the four-loop result for the $\rho$ parameter in the $\MSbar$ scheme
\begin{equation}
  \delta\rho_t^{\mbox{\tiny{(4 loops)}}}=3{G_F
    \overline{m}_t^2\over8\sqrt{2}\pi^2}\left({\alpha_s\over\pi}\right)^3
  \underbrace{(-3.2866+1.6067)}_{-1.6799}\,.
\end{equation}
The first term has been evaluated in Ref.~\cite{Schroder:2005db},
the second one, which (in the $\MSbar$ scheme) leads 
to a reduction by a factor of about $1/2$, 
in Refs.~\cite{Chetyrkin:2006bj,Boughezal:2006xk}. 
For the numerical evaluation the translation from the
$\MSbar$ to the pole mass is more convenient and one finds
\begin{equation}
  \delta\rho_t^{\mbox{\tiny{(4 loops)}}} = 3{G_F
    M_t^2\over8\sqrt{2}\pi^2}
    \left(\frac{\alpha_s}{\pi}\right)^3 (-93.1501) 
\end{equation}
which corresponds to a shift in $M_W$ by about 2~MeV, similar to
the three loop term of order $\alpha_s X_t^2$.


\section{Charm- and bottom-quark masses\label{sec:III}}

The precise determination of the charm- and bottom-quark masses from
relativistic four-loop moments can be considered as one of the truly
remarkable successes of quantum field theory with remarkable agreement
between perturbative~\cite{Kuhn:2007vp,Chetyrkin:2009fv} and lattice
methods~\cite{Allison:2008xk,McNeile:2010ji,Colquhoun:2014ica}. Let us
first give a more detailed motivation, then present the theoretical
framework and finally compare the results based on perturbative and
lattice methods.

Precise bottom quark masses
enter many $B$ physics quantities.  During the past years
significant progress has been made on the one hand in the analysis of
$B$-meson decays and, on the other hand, the $\Upsilon$ spectroscopy.  In
particular the latter has led to a fairly consistent result of
$m_b(m_b)=4.193^{+0.022}_{-0.035}$~GeV~\cite{Beneke:2014pta} (see also
Ref.~\cite{Penin:2014zaa} where $m_b(m_b)=4.169\pm0.009$~GeV has been
obtained) in excellent agreement with the result based on sum rules discussed
in detail below.

Let us further motivate the need for precise quark masses for the case of the
bottom-quark mass, which enters a number of physical observables.  Most
prominently, we want to mention the decay of a Higgs boson
into bottom quarks which, using the
scalar correlator to five-loop precision~\cite{Baikov:2005rw}, can be written
in the form
\begin{eqnarray}
  \Gamma(H\to b\bar{b})&=&{G_F M_H^2\over4\sqrt{2}\pi}m_b^2(M_H)\*R^{(S)}(M_H)\\
  R^{(S)}(M_H)&=&
  1
  + 5.667  \left({\alpha_s \over \pi}\right) 
  + 29.147 \left({\alpha_s \over \pi}\right)^2 \nonumber \\
  &&\mbox{}+41.758 \left({\alpha_s \over \pi}\right)^3 
  - 825.7  \left({\alpha_s \over \pi}\right)^4 \nonumber\\
  &=& 
  1 
  + 0.19551 
  + 0.03469 \nonumber\\
  &&\mbox{}+0.00171 
  - 0.00117.
\end{eqnarray}
The theory uncertainty, which is generously taken from a variation of the
scale parameter between $M_H/3$ and $3M_H$, is reduced from $5\permil$ for the
four-loop to $1.5\permil$ for the five-loop result.  Thus, the main
uncertainty is induced from the uncertainty of the bottom quark mass which (at
energy scale 10~GeV) is given by~\cite{Chetyrkin:2009fv}
\[
\hspace*{-0.5cm}
m_b(10~\mbox{GeV})=\left(3610-{\alpha_s-0.1189\over0.002}\cdot12\pm11\right)~\mbox{MeV}
\,.
\]
The running from 10~GeV to $M_H$ depends on the anomalous mass
dimension $\gamma_m$, the $\beta$ function and on $\alpha_s$. With the present
knowledge (i.e. four-loop anomalous dimensions as implemented in {\tt
  RunDec}~\cite{Chetyrkin:2000yt,Schmidt:2012az}) one finds
\begin{eqnarray}
  m_b(M_H)=2759\pm\left.8\right|_{m_b}\pm\left.27\right|_{\alpha_s}~\mbox{MeV}
  \,.
  \label{eq::mbMH}
\end{eqnarray}

It is interesting to investigate the effect of the five-loop anomalous
dimensions. Taking into account the 
$\gamma_m$ to five-loop accuracy~\cite{Baikov:2014pja,Baikov:2014qja}
together with $\beta_4$, the five-loop contribution to the $\beta$
function of QCD, which is still unknown, one obtains the following
uncertainties
\begin{eqnarray}
  \delta m_b^2(M_H)\over m_b^2(M_H)&=& 
  -1.3\times10^{-4} (\beta_4/\beta_0=0)\nonumber\\[-0.25cm]
  &=& -4.3\times10^{-4} (\beta_4/\beta_0=100)\nonumber\\ 
  &=& -7.3\times10^{-4} (\beta_4/\beta_0=200)\,,
\end{eqnarray}
which lead to an uncertainty of a few MeV in $m_b(M_H)$
which is small compared to the current error shown in Eq.~(\ref{eq::mbMH}).

Another motivation which also points to a precision around 10~MeV is
based on the picture of Yukawa unification. In this approach 
$\lambda_\tau\sim\lambda_b\sim\lambda_t$ at the GUT scale. In effect
this implies $\delta m_b/m_b\sim\delta m_t/m_t$. Assuming a precision
of the top-quark mass $\delta m_t\approx0.5$~GeV then leads to a precision
of $\delta m_b\approx10$~MeV, consistent with our finding below.

\subsection*{SVZ sum rules, moments and tadpoles}

\begin{figure}[!h]
  \begin{center}
    \includegraphics[angle=-1,width=7cm,angle=180]{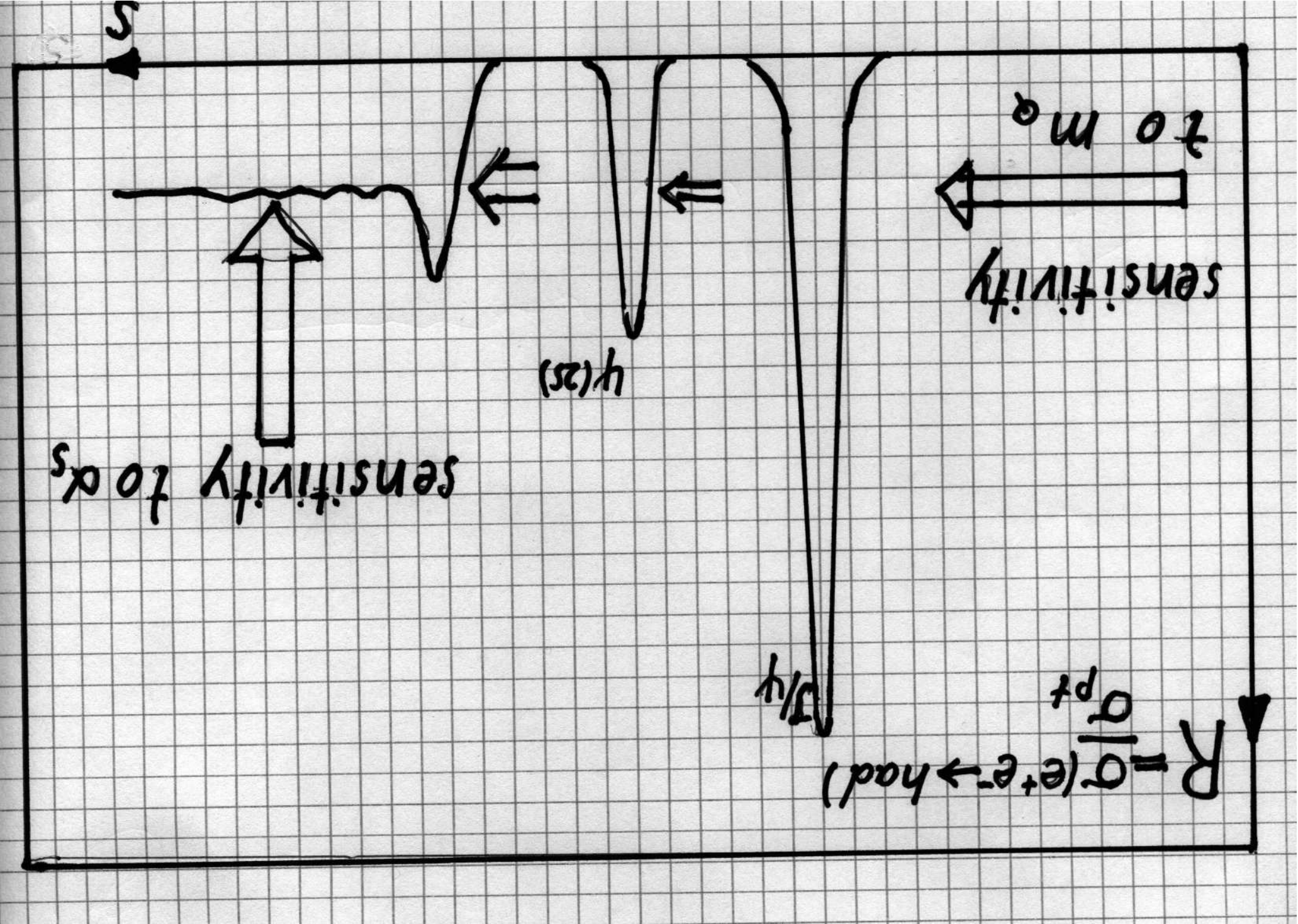}
  \end{center}
  \vspace*{-0.6cm}
  \caption{Sketch of the $R$ ratio in the charm-threshold region.\label{fig:r}}
\end{figure}

The main idea originally advocated in Ref.~\cite{Novikov:1977dq} is
based on the observation (cf. Fig.~\ref{fig:r}), that the cross
section for hidden ($J/\Psi$ plus $\Psi(2S)$) plus open
($D\overline{D}$ plus resonances) charm production is well
described by perturbation theory, if the analysis is restricted to
sufficiently low moments.

Let us first recall some definitions. The
two-point correlation function
\begin{eqnarray*}
  \!\!\!\!(-q^2g_{\mu\nu}+q_{\mu}q_{\nu})\Pi(q^2)
  =\ri\!\int\!dx e^{iqx} \langle0| T j_{\mu}(x)j_{\nu}(0)|0\rangle
\end{eqnarray*}
is related to the electromagnetic current $j_\mu$ as follows
\begin{equation}
  R(s)=12\pi\mbox{Im}\left[\Pi(q^2=s+\ri\varep)\right]\,.
\end{equation}
In fact, we are only interested in lowest moments of 
$\Pi$, corresponding  to the first few terms of the
Taylor expansion of $\Pi(q^2)$:
\begin{equation}
  \Pi(q^2)=Q_q^2 {3\over16\pi^2}\sum_{n\ge0} \overline{C}_n z^n\,,
\end{equation}
where $Q_q$ corresponds to the charge of the considered quark.
Here $z=q^2/(4m_q^2)$ and $m_q=m_q(\mu=m_q)$ is the $\MSbar$ mass at
the scale $\mu=m_q$. Let us, for definiteness, restrict the
following discussion to the charm quark, i.e., $q=c$.

\begin{figure}[!h]
\begin{center}
\includegraphics[width=1.5cm]{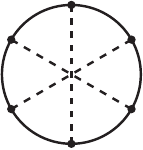}\hspace{0.5ex}
\includegraphics[width=1.5cm]{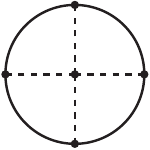}\hspace{0.5ex}
\includegraphics[width=1.5cm]{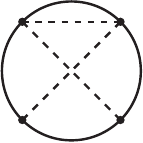}\hspace{0.5ex}
\includegraphics[width=1.5cm]{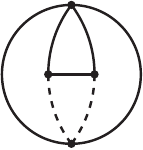}\\
\includegraphics[width=1.5cm]{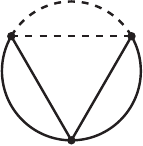}\hspace{0.5ex}
\includegraphics[width=1.5cm]{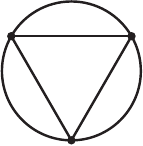}\hspace{0.5ex}
\includegraphics[width=1.5cm]{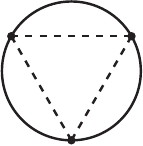}\hspace{0.5ex}
\includegraphics[width=1.5cm]{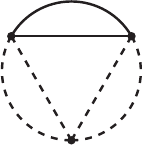}\\
\includegraphics[width=1.5cm]{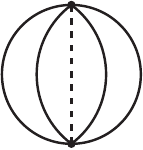}\hspace{0.5ex}
\includegraphics[width=1.5cm]{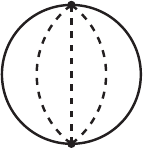}\hspace{0.5ex}
\includegraphics[width=1.5cm]{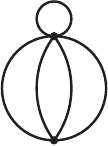}\hspace{0.5ex}
\includegraphics[width=1.5cm]{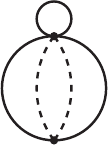}\\
\includegraphics[width=1.5cm]{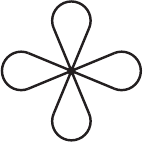}
\end{center}
\vspace*{-0.6cm}
\caption{The 13 master integrals. Solid lines denote massive
  propagators; dashed lines represent massless propagators.\label{fig:MI}}
\end{figure}

For the moments one finds 
\begin{eqnarray}
\overline{C}_n&=&\overline{C}_n^{(0)}
               + \frac{\alpha_\mathrm{s}}\pi \overline{C}_n^{(1)}
               + \left( \frac{\alpha_\mathrm{s}}\pi\right)^2 \overline{C}_n^{(2)}\nonumber\\
             &&+ \left( \frac{\alpha_\mathrm{s}}\pi\right)^3 \overline{C}_n^{(3)}
               + \dots,
\end{eqnarray}
if the renormalization scale is set to $\mu=m_q$ or
\begin{eqnarray}
 \overline{C}_n &=&  \overline{C}_n^{(0)}
                   + \frac{\alpha_\mathrm{s}}{\pi} \big(
                          \overline{C}_n^{(10)}
                       +  \overline{C}_n^{(11)}l_{m_c}
                                                \big)
\nonumber\\&&\mbox{}
                 +\left( \frac{\alpha_\mathrm{s}}{\pi}\right)^2 \big(
                          \overline{C}_n^{(20)}
                       +  \overline{C}_n^{(21)}l_{m_c}
                       +  \overline{C}_n^{(22)}l_{m_c}^2       \big)
\nonumber\\&&\mbox{}
      +\left( \frac{\alpha_\mathrm{s}}\pi\right)^3 
                                            \big(
              \overline{C}_n^{(30)} 
            +  \overline{C}_n^{(31)}l_{m_c}
            +  \overline{C}_n^{(32)}l_{m_c}^2
\nonumber\\&&\mbox{}\qquad\quad 
            +  \overline{C}_n^{(33)}l_{m_c}^3\big)
            + \dots,
\end{eqnarray}
if one is interested in the generic form with \mbox{$l_{m_c} = \ln
  (m_c^2(\mu)/\mu^2)$}.  The next-to-next-to-leading order calculation had
been performed already nearly twenty years
ago~\cite{Chetyrkin:1995ii,Chetyrkin:1996cf,Chetyrkin:1997mb} and is available
for all four (vector, axial, scalar and pseudoscalar) correlators. The
original evaluation was up to $n=8$. More recently, this has been extended to
$n=30$~\cite{Boughezal:2006uu,Maier:2007yn}. Now this project has been pushed
to N$^{3}$LO, and the results will be described in the following.  In a first
step the $n_f^2$ contribution has been computed for $\overline{C}_0$ and
$\overline{C}_1$~\cite{Chetyrkin:2004fq}, then the complete result became
available. The reduction of the many different diagrams has been performed to
13 master integrals, shown in Fig.~\ref{fig:MI}, using the Laporta
algorithm~\cite{Laporta:2001dd}.  Subsequently these 13 remaining integrals
are evaluated, using originally a combination of numerical and analytical, now
purely analytical methods.

The reduction of hundreds of integrals to master integrals has
been achieved, originally for $\overline{C}_0$ and
$\overline{C}_1$~\cite{Chetyrkin:2006xg,Boughezal:2006px},
subsequently for $\overline{C}_2$ and $\overline{C}_3$, using the
program Crusher~\cite{Maier:2008he,Maier:2009fz}. In the meantime all
master integrals are known in closed  analytic
form to high order in $\varepsilon$, using results by a number of
different authors (see
Refs.~\cite{Laporta:2002pg,Schroder:2005va,Chetyrkin:2006dh,Lee:2010hs} and
references therein).  The results for $\overline{C}_4$ up to
$\overline{C}_{10}$ are known approximately, using an approximation
based on additional information from low energies ($q^2=0$), from
threshold ($q^2=4m^2$) and from high
energy~\cite{Hoang:2008qy,Kiyo:2009gb,Greynat:2010kx}. Closely related
results are also known for axial, scalar and pseudoscalar
correlators~\cite{Sturm:2008eb,Maier:2009fz,Kiyo:2009gb}. These can be
used for the investigation of correlators on the
lattice~\cite{Allison:2008xk} and will not be investigated further in
this more phenomenological study.
The heavy quark vector current correlator for $q^2\ll m^2$
has also been determined in the large $\beta_0$
limit~\cite{Grozin:2004ez} in order to study the
large-order behaviour.

The moments are directly related to measurements as follows. From
theoretical considerations one finds
\begin{eqnarray}
  \mathcal{M}_n^{\mbox{\tiny{th}}}&\equiv&\left.{12\pi^2\over n!}
    \left({d\over dq^2}\right)^n\Pi_c(q^2)\right|_{q^2=0}
  \nonumber\\ 
  &=&
  {9\over4}Q_c^2\left({1\over4m_c^2}\right)^n\overline{C}_n\,,
\end{eqnarray}
where the quantity $\overline{C}_n$ depends on $\alpha_s(\mu^2)$
and $\ln(m_c^2/\mu^2)$. As default value for $\mu$ we
use $\mu=3$~GeV.

To obtain experimental moments one considers the correlator
given by
\begin{equation}
  \Pi_c(q^2)={q^2\over12\pi^2}\int\D s {R_c(s)\over s(s-q^2)}+ \mbox{subtraction}\,,
\end{equation}
which leads to
\begin{equation}
  \mathcal{M}^{\mbox{\tiny{exp}}}_n=\int{\D s\over s^{n+1}} R_c(s)\,.
  \label{eq:Mnexp}
\end{equation}
Imposing the constraint
$\mathcal{M}_n^{\mbox{\tiny{exp}}}=\mathcal{M}_n^{\mbox{\tiny{th}}}$
leads to $m_c$ at the scale $\mu=3$~GeV, a result that could be
easily translated to arbitrary values of $\mu$.

\begin{table}[t]
\begin{center}
\scalebox{0.7}{
\begin{tabular}{l|lll|l||l}
\hline
$n$ & ${\mathcal{M}_n^{\rm res}}$
& ${\mathcal{M}_n^{\rm thresh}}$
& ${\mathcal{M}_n^{\rm cont}}$
& ${\mathcal{M}_n^{\rm exp}}$
& ${\mathcal{M}_n^{\rm np}}$
\\
& $\times 10^{(n-1)}$
& $\times 10^{(n-1)}$
& $\times 10^{(n-1)}$
& $\times 10^{(n-1)}$
& $\times 10^{(n-1)}$
\\
\hline
$1$&$  0.1201(25)$ &$  0.0318(15)$ &$  0.0646(11)$ &$  0.2166(31)$ &$ -0.0001(2)$ \\
$2$&$  0.1176(25)$ &$  0.0178(8)$ &$  0.0144(3)$ &$  0.1497(27)$ &$ 0.0000(0)$ \\
$3$&$  0.1169(26)$ &$  0.0101(5)$ &$  0.0042(1)$ &$  0.1312(27)$ &$  0.0007(14)$ \\
$4$&$  0.1177(27)$ &$  0.0058(3)$ &$  0.0014(0)$ &$  0.1249(27)$ &$  0.0027(54)$ \\
\hline
\end{tabular}
}
\caption{
  \label{tab:Mexp} The experimental moments in $(\mbox{GeV})^{-2n}$ as defined in
  Eq.~(\ref{eq:Mnexp}) are shown, separated according to the
  contributions from the narrow resonances, the charm threshold region
  and the continuum region above $\sqrt{s}=4.8$~GeV. The last column
  gives the contribution from the gluon condensate.}
\end{center}
\end{table}

\begin{figure}[tb]
  \begin{center}
    \includegraphics[width=\columnwidth]{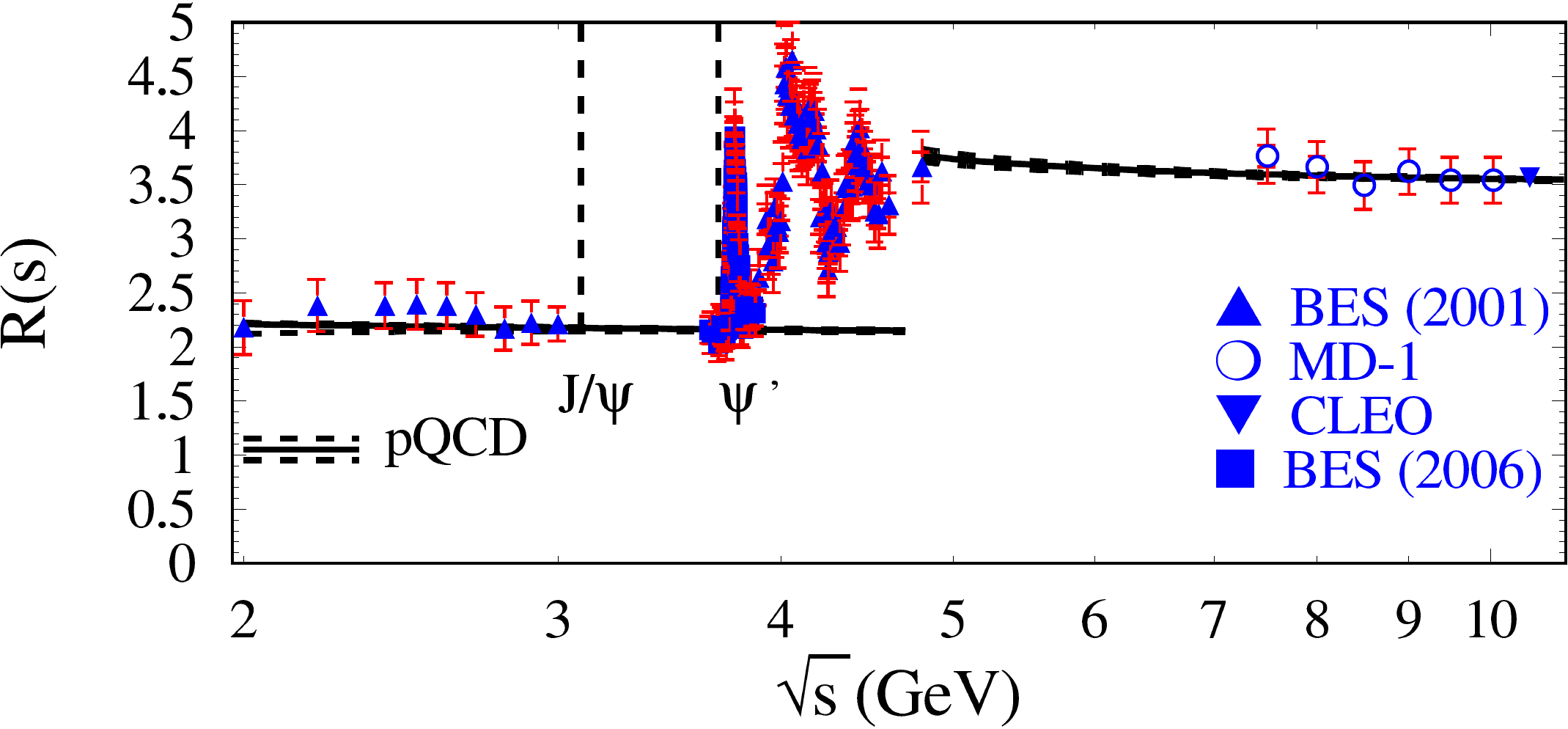}
  \end{center}
  \vspace*{-0.6cm}
  \caption{The normalized cross section $R(s)$ between 2~GeV and 10~GeV.
    The solid line corresponds to the theoretical prediction. The
    uncertainties obtained from the variation of the input parameters and
    of $\mu$ are indicated by the dashed curves. The inner and outer error
    bars give the statistical and systematical uncertainty,
    respectively. The data points are from
    BES~\cite{Bai:2001ct,Ablikim:2006mb}, MD-1~\cite{Blinov:1993fw} and
    CLEO~\cite{Ammar:1997sk}. The vertical  dashed lines correspond to
    the location of the $J/\Psi$ and $\Psi'$ resonances.\label{fig:tbd}}
\end{figure}

Let us first discuss the ingredients for charm, and then for bottom
quarks. The results for the electronic widths of $J/\Psi$ and $\Psi'$
are taken from the combination of BES, CLEO and BABAR experiments,
and for the continuum $R(s)$ from BES. For
the charm case there is also a non-perturbative contribution
which is, however, negligible for the three lowest moments and
remains relatively small even for the fourth. A careful investigation
of non-perturbative terms combined with the extrapolation of $R_{uds}$
as well as $R_c$ in the region sufficiently far above the respective
threshold leads to a remarkably consistent result, with errors on
$m_c(3~\mbox{GeV})$, as extracted for $n=1$ to $3$, below
20~MeV. The result for the moments individually split into
contributions from the narrow resonances, threshold and continuum
region is shown in Tab.~\ref{tab:Mexp}.  $R(s)$ around the charm
threshold region is shown in Fig~\ref{fig:tbd}.

\begin{figure}[!h]
  \begin{center}
    \includegraphics[width=\columnwidth]{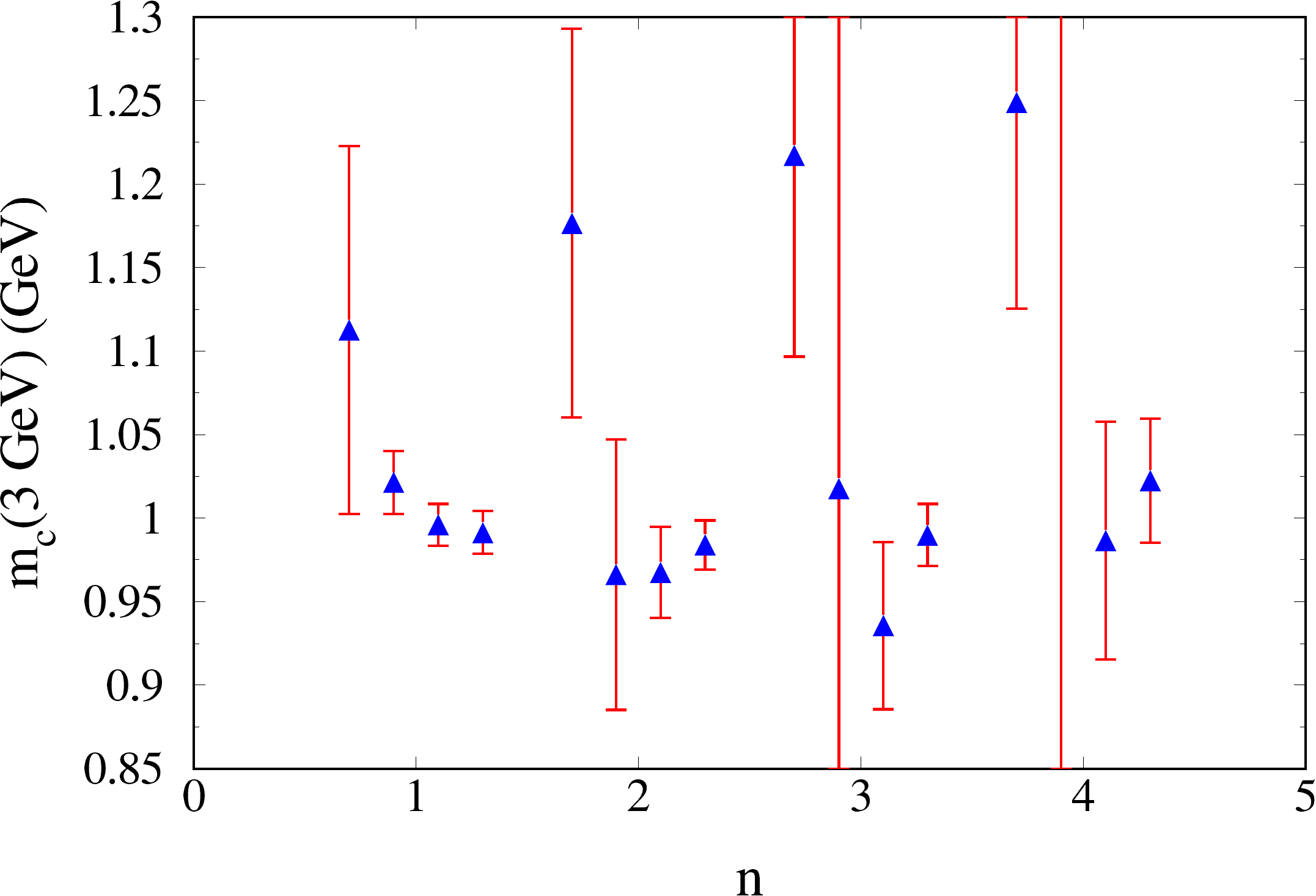}
  \end{center}
  \vspace*{-0.6cm}
  \caption{Dependence of $m_c(3~\mbox{GeV})$ on the number of moments $n$ and on $\mathcal{O}\left(\alpha_s^i\right)$ for $i=0,\dots,3$.
    \label{fig:mcmoments}}
\end{figure}

In particular we observe a remarkable consistency between the results
for $n=1,2,3$ and $4$ and a relatively small shift when moving from
three- to four-loop approximation (cf. Fig.~\ref{fig:mcmoments}). 
\begin{table}[!h]
\begin{center}
\begin{tabular}{l|r@{}l|rrrr|c}
\hline
$n$ & \multicolumn{2}{|c|}{$m_c(3~\mbox{GeV})$} & exp & $\alpha_s$ & $\mu$ & np & 
total
\\
\hline
       1&   &986&  9&   9&   2&  1 & 13 \\
       2&   &976&  6&  14&   5&  0 & 16 \\
       3&   &978&  5&  15&   7&  2 & 17 \\
       4&  1&004&  3&   9&  31&  7 & 33 \\
\hline
\end{tabular}
\caption{\label{tab:mc1}The second column shows the results for
  $m_c(3~\mbox{GeV})$ in MeV. The errors in the four inner columns are
  from experiment, $\alpha_s$, variation of $\mu$
  and the gluon condensate. The last column shows the total error.
}
\end{center}
\end{table}

In taking the lowest moment as our
final result we find~\cite{Chetyrkin:2009fv}
\begin{equation}
  m_c(3~\mbox{GeV})=986\pm13~\mbox{MeV}.
\end{equation}
When converted from $\mu=3$~GeV to the scale $m_c$, this is modified to
$m_c(m_c)=1279\pm13$~MeV, nicely consistent with other
determinations~\cite{Allison:2008xk,McNeile:2010ji}. The
robustness of our result is demonstrated in Fig.~\ref{fig:mcmoments}, where the
results are compared for different orders $\mathcal{O}(\alpha_s^i)$,
with $i$=0, 1, 2 and 3 and for different moments, with $n$ varying
between $n=1$ and $n=4$. 

\begin{figure}[!h]
  \begin{center}
    \includegraphics[width=\columnwidth]{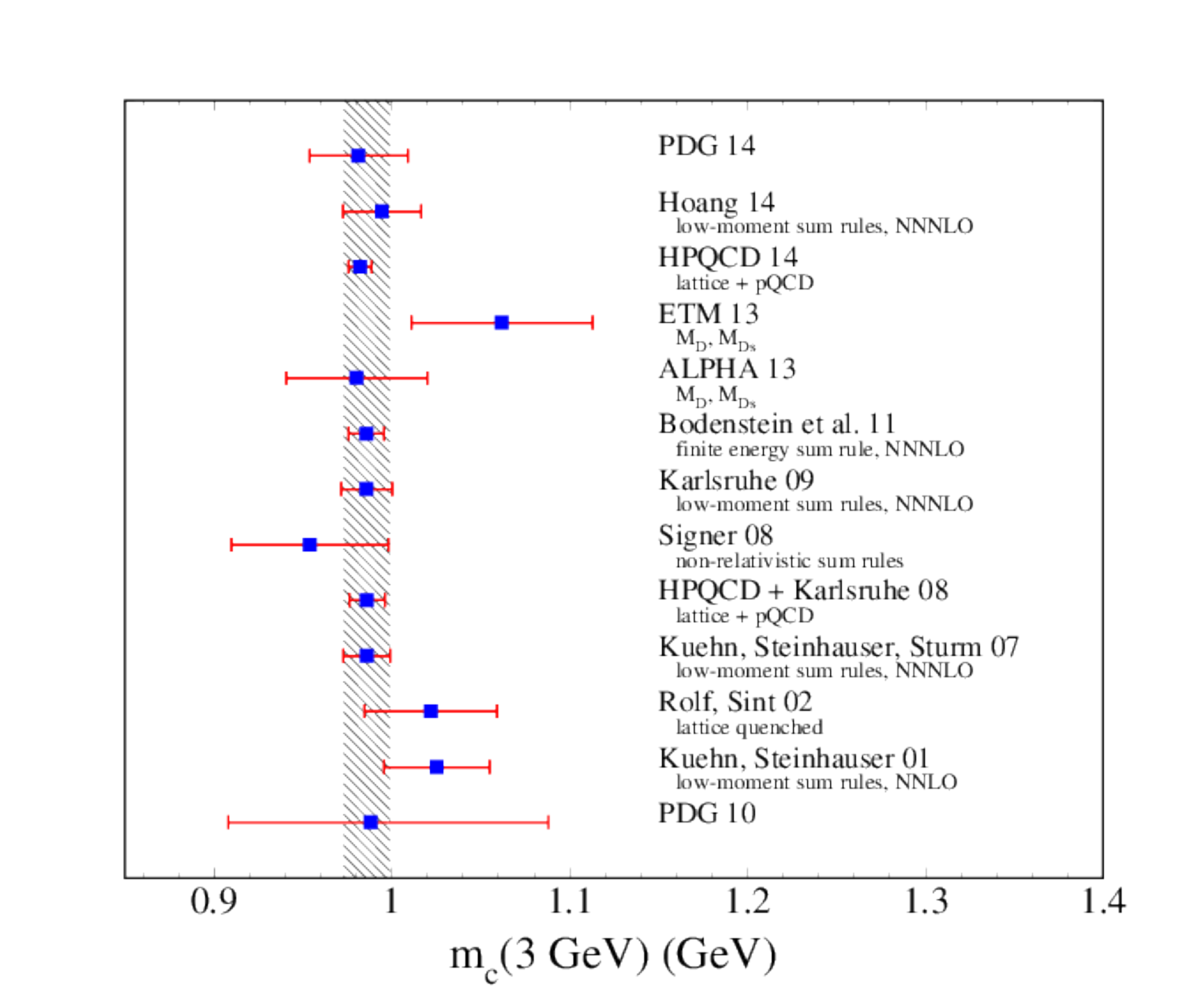}
  \end{center}
  \vspace*{-0.6cm}
  \caption{Comparison of $m_c(3~\mbox{GeV})$ with several other
    results.\label{fig:mccomparison}}
\end{figure}

The result can be compared to those from a large
number of results, which are based on various different
observables. Fig~\ref{fig:mccomparison} shows a compilation 
of recent analyses~\cite{Kuhn:2001dm,Rolf:2002gu,Kuhn:2007vp,Allison:2008xk,Signer:2008da,Chetyrkin:2009fv,Chetyrkin:2010ic,Bodenstein:2011ma,Heitger:2013oaa,Carrasco:2014cwa,Chakraborty:2014aca,Dehnadi:2014kya,Agashe:2014kda}. 

Similar considerations are applicable for the corresponding
investigations of the $\Upsilon$-resonances and the mass of the bottom
quark. For convenience of the reader we again list in
Tab.~\ref{tab::mb_mom} separately the contributions from the narrow
resonances ($\Upsilon(1S)-\Upsilon(4S)$)~\cite{Yao:2006px}, the
threshold region (10.618~GeV-11.2~GeV)~\cite{Aubert:2008ab} and the
perturbative continuum ($E>11.2$~GeV).

\begin{table}[!h]
  \centering
{\scalefont{0.8}
  \begin{tabular}{c|lll|l}
\hline
$n$&${\cal M}_n^{\mbox{res,(1S-4S)}}$&${\cal M}_n^{\mbox{thresh}}$&${\cal M}_n^{\mbox{cont}}$&${\cal M}_n^{\mbox{exp}}$
\\
& $\times 10^{(2n+1)}$
& $\times 10^{(2n+1)}$
& $\times 10^{(2n+1)}$
& $\times 10^{(2n+1)}$\\
\hline
    $1$&$   1.394(23)$ &$   0.287(12)$ &$   2.911(18)$ &$   4.592(31)$ \\
$2$&$   1.459(23)$ &$   0.240(10)$ &$   1.173(11)$ &$   2.872(28)$ \\
$3$&$   1.538(24)$ &$   0.200(8)$ &$   0.624(7)$ &$   2.362(26)$ \\
$4$&$   1.630(25)$ &$   0.168(7)$ &$   0.372(5)$ &$   2.170(26)$ \\\hline
  \end{tabular}
  \caption{Moments for the bottom quark system in $(\mbox{GeV})^{-2n}$}
  \label{tab::mb_mom}
}
\end{table}

For the lowest moment the latter gives the
main contribution, starting from the second moment the resonance and the
threshold regions are again dominant. In particular moments number two
and three offer a fair compromise between smallness of the error and the
contribution of
the threshold region. Significant progress has been made between the
first measurements of CLEO~\cite{Besson:1984bd} and the more recent one
by the BABAR collaboration in particular in the continuum region. 
As expected~\cite{Kuhn:2007vp} the original
CLEO result is too large by a factor around 1.3 but reproduces well the
qualitative behaviour. The more recent BABAR
result~\cite{Aubert:2008ab}, however, is significantly more precise and
well in agreement with the expectations for the continuum, based on the
parton  cross section. Let us note in this connection that the
original BABAR result for $R_b(s)$ has to be deconvoluted with respect
to initial state radiation (ISR), a fact that leads to a slight shift of
$R_b(s)$.

The results for the bottom quark mass
for the lowest four moments are given in
Tab.~\ref{tab:mb}~\cite{Chetyrkin:2009fv,Chetyrkin:2010ic}.
For our final results we choose $n=2$ which leads to
\begin{eqnarray}
  m_b(m_b) &=&\,4.163 \pm 0.016~\mbox{GeV}\,,\nonumber\\
  m_b(10~\mbox{GeV}) &=&\,3.610 \pm 0.016~\mbox{GeV}\,, \nonumber\\
  m_b(M_H)&=&2.759 \pm 0.028~\mbox{GeV}\,.
\label{eq:mb}
\end{eqnarray}

\begin{table}[!h]
\begin{center}
\begin{tabular}{l|l|rrr|l|l}
\hline
$n$ & $m_b(10~\mbox{GeV})$ & 
exp & $\alpha_s$ & $\mu$ &
total &$m_b(m_b)$
\\
\hline
        1&  3597&  14&  7&   2&  16&  4151 \\
        2&  3610&  10&  12&  3&  16&  4163 \\
        3&  3619&  8&  14&   6&  18&  4172 \\
        4&  3631&  6&  15&  20&  26&  4183 \\
\hline
\end{tabular}
\end{center}
\vspace*{-0.6cm}
\caption{The different columns show the results for
  $m_b(10~\mbox{GeV})$ in the second column, obtained from the
  different moments listed in the first column. The last column gives
  the value of $m_b(m_b)$. The three inner columns give the
  uncertainty due to the error in the experimental moments (exp), the
  uncertainty due to the error in $\alpha_s$ and the uncertainty due
  to the residual scale dependence $\mu$. The second to last column
  gives the total uncertainty. All masses and uncertainties are in
  units of MeV.\label{tab:mb}}
\end{table}

\begin{figure}[!h]
  \begin{center}
    \includegraphics[width=\columnwidth]{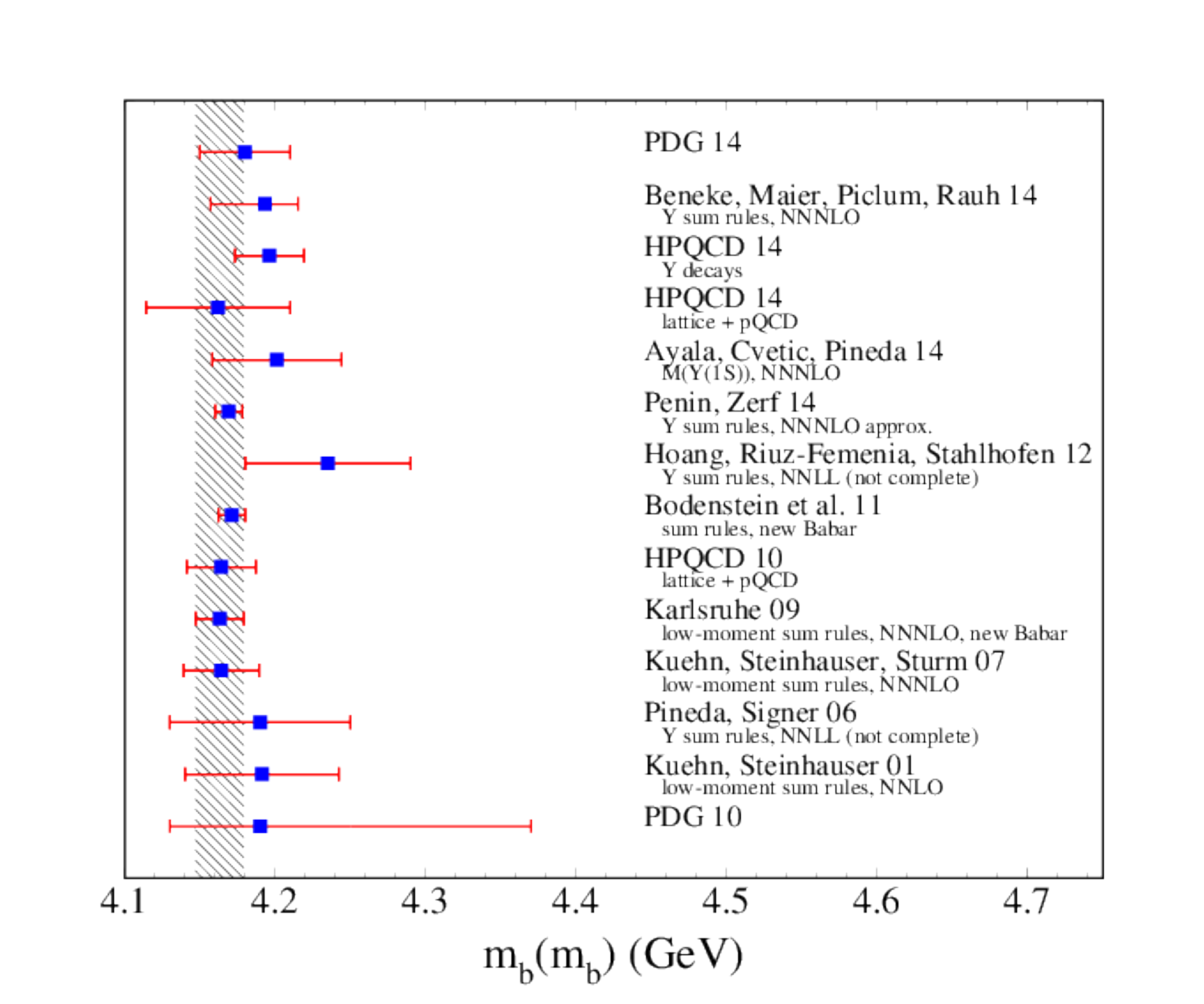}
  \end{center}
  \vspace*{-0.6cm}
  \caption{Comparison of $m_b(m_b)$ with several other
    determinations.\label{fig:mbcomparison}}
\end{figure}

The consistency, when comparing the results for the lowest four
moments is very close to the one from 2007~\cite{Kuhn:2007vp}, where
only estimates were available for the four-loop term of $n=2, 3$ and
$4$. Furthermore, only recalibrated results for the continuum
corresponding to the aforementioned factor 1.3 were available.  
The result for
$m_b(m_b)$ can also be compared to those from other studies in
Fig.~\ref{fig:mbcomparison}~\cite{Kuhn:2001dm,Pineda:2006gx,Kuhn:2007vp,Chetyrkin:2009fv,Chetyrkin:2010ic,McNeile:2010ji,Bodenstein:2011fv,Hoang:2012us,Penin:2014zaa,Chakraborty:2014aca,Colquhoun:2014ica,Ayala:2014yxa,Beneke:2014pta,Agashe:2014kda}. Although somewhat towards the low side,
the result are well compatible with those of earlier investigations.

\begin{figure}[htb]
  \begin{center}
    \includegraphics[width=\columnwidth]{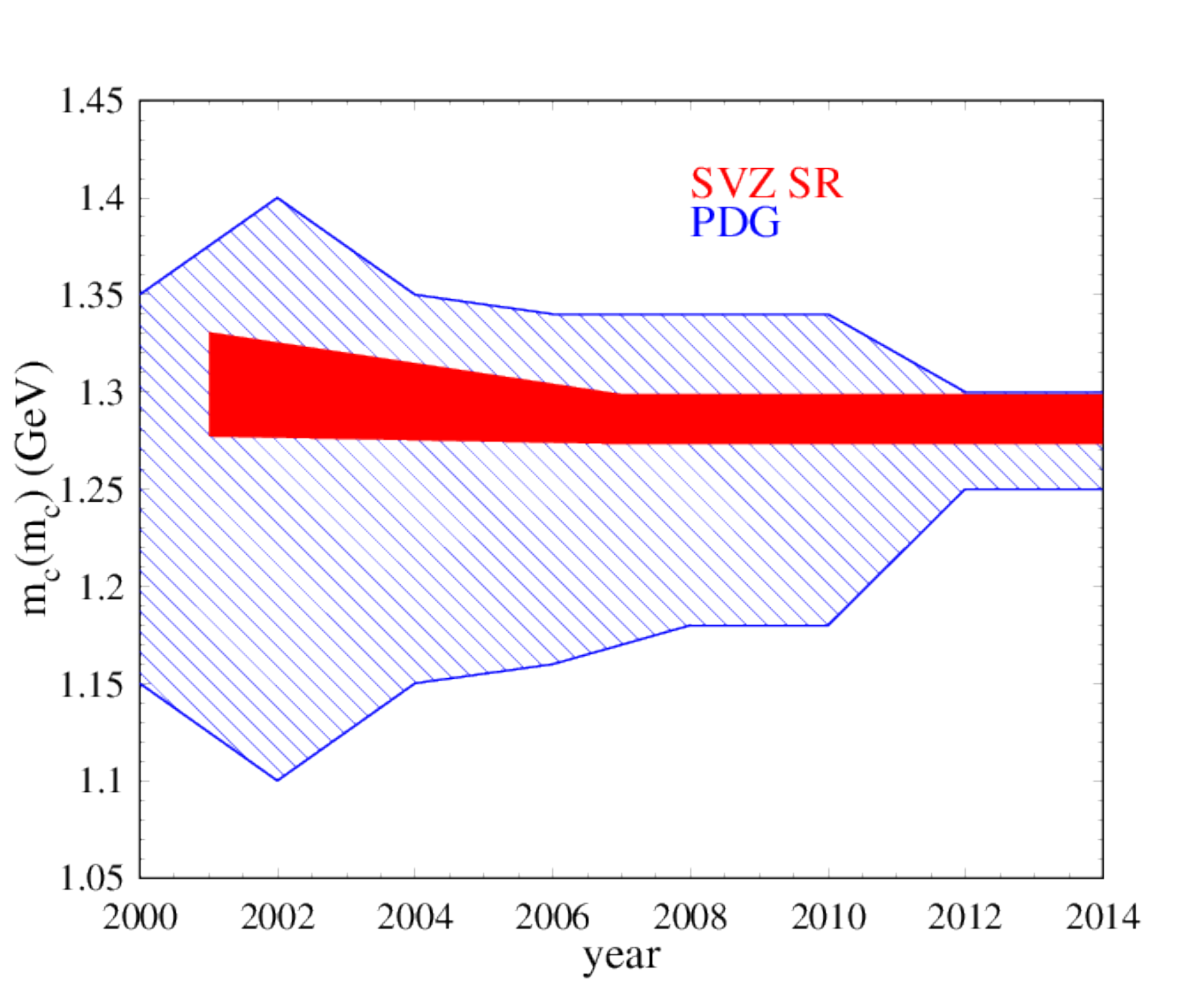}
    \includegraphics[width=\columnwidth]{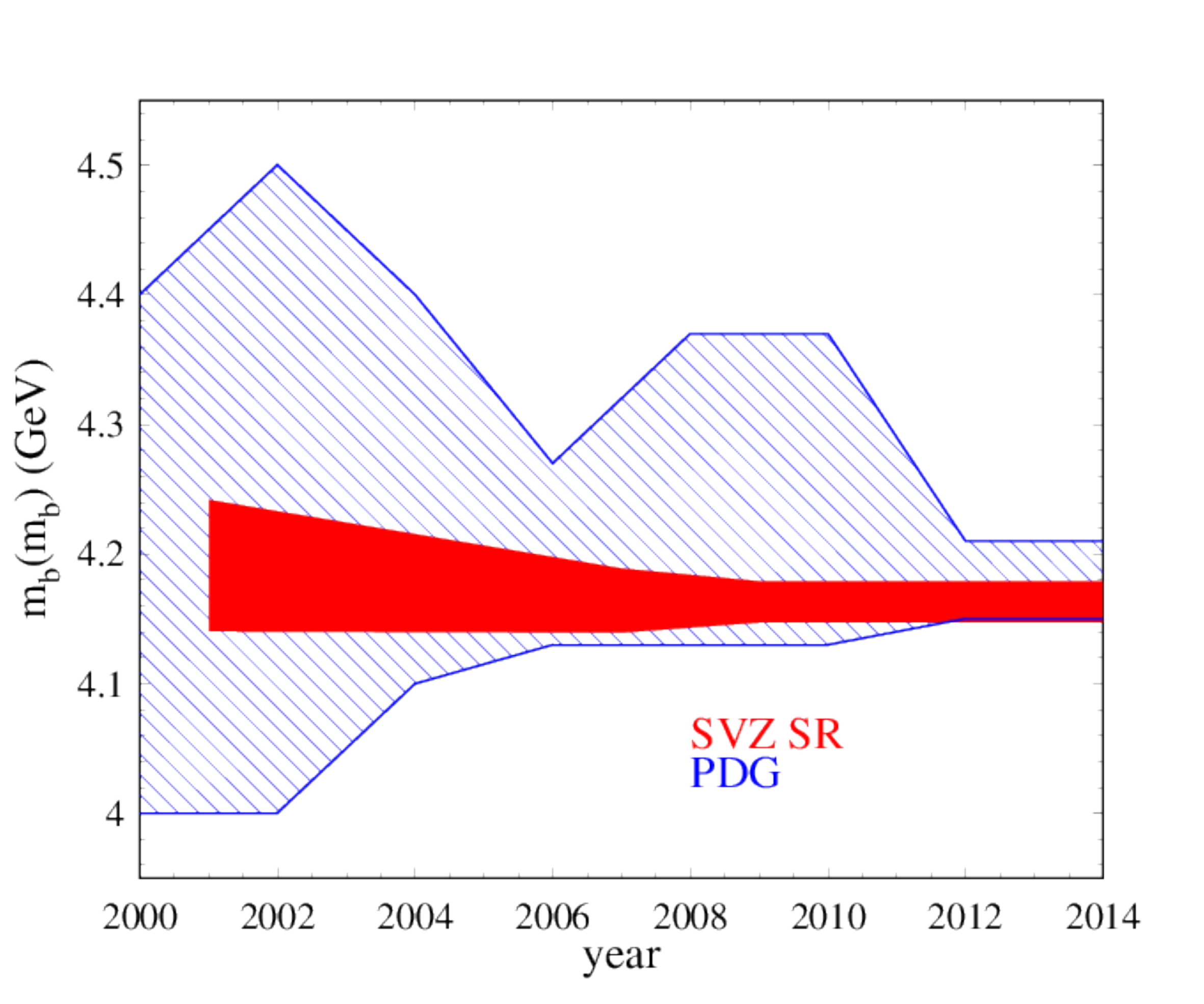}
  \end{center}
  \vspace*{-0.6cm}
  \caption{\label{fig::ka_pdg}
    Comparison of $m_c(m_c)$ and $m_b(m_b)$ as obtained using
    low-moment sum
    rules~\cite{Kuhn:2001dm,Kuhn:2007vp,Chetyrkin:2009fv,Chetyrkin:2010ic}
    (narrow band) and the values from the PDG between 2000 and 2014.}
\end{figure}

In Fig.~\ref{fig::ka_pdg} the results for the charm and bottom quark masses as
obtained from the low-moment sum
rules~\cite{Kuhn:2001dm,Kuhn:2007vp,Chetyrkin:2009fv,Chetyrkin:2010ic} are
compared with the numerical values proposed by the PDG for the years between
2000 and 2014. It is interesting to note that the extracted mass values, which
were first based on three, later on four-loop perturbative input,
remained rather constant whereas the PDG numbers seem to converge towards
these results.

\section{Further applications of massive tadpoles\label{sec:IV}}

\subsection*{Decoupling function at four loops}

In many QCD applications the mass of a heavy quark $m$ is much larger
than the characteristic momentum scale~$\sqrt{s}$ of a considered
physical process.  As a result these different mass scales involved in
the process can lead to potentially large logarithms like
$\log(\mu/\sqrt{s})$ or $\log(\mu/m)$ when using an MS-like
renormalization scheme. In such a situation one can not set the
renormalization scale $\mu$ to two different mass scales
simultaneously, so that a proper choice of $\mu$ in order to avoid
large logarithms is not possible anymore. However, by ``integrating
out'' the heavy quark field one can construct an effective field
theory with $n_l=n_f-1$ light quarks only, where $n_f$ is the number
of quark flavours.

The $\MSbar$ coupling constants $\alpha_s^{(n_f)}$ and
$\alpha_s^{(n_l)}$ of the quark-gluon interaction in the full
$n_f$-flavor QCD and the effective $n_l$-flavor one are different and
are related by the decoupling function
$\zeta_g(\mu,\alpha_s^{(n_f)}(\mu),m)$ through the matching condition
\begin{equation}
  \alpha_s^{(n_l)}(\mu) =
  \zeta_g^2(\mu,\alpha_s^{(n_f)}(\mu),m)\,\,\alpha_s^{(n_f)}(\mu).
\end{equation}
At leading order the decoupling function is equal to one, but receives
corrections in higher orders of perturbation theory.  This matching
condition for the $\MSbar$ strong coupling constant $\alpha_s$ at a
heavy quark threshold has been computed in
Refs.~\cite{Chetyrkin:2005ia,Schroder:2005hy} to four-loop order.  The
decoupling function can be determined through the computation of
polarization functions. The bare relation for $\zeta_g^0$
reads~\cite{Chetyrkin:1997un} 
\begin{equation}
\zeta_g^0={\tilde{\zeta}_{1}^{0}\over\tilde{\zeta}_{3}^{0}\sqrt{\zeta_{3}^{0}}},
\end{equation}
where
\begin{eqnarray}
\zeta_{3}^{0}&=&1+\Pi_{G}^{0h}(0),\nonumber\\
\tilde{\zeta}_{3}^{0}&=&1+\Pi_{c}^{0h}(0),\nonumber\\
\tilde{\zeta}_{1}^{0}&=&1+\Gamma_{G\bar{c}c}^{0h}(0,0)
\end{eqnarray}
with the gluon $G$ and ghost $c$ vacuum polarization functions
$\Pi_G^{0h}(q^2)$ and $\Pi_{c}^{0h}(q^2)$. The vertex function
$\Gamma_{G\bar{c}c}^{0h}(p,k)$ is the one-particle irreducible part of
the amputated Green's function, where $p$ and $k$ are the outgoing
four momenta of the fields $c$ and $G$, respectively.  The computation
of these functions leads again to the evaluation of tadpole diagrams
up to four-loop order. The four-loop contribution can be expressed in
terms of the 13 master integrals shown in Fig.~\ref{fig:MI}. The
renormalized decoupling function is obtained from
\begin{equation}
\alpha_s^{(n_l)}=\left(
{Z_g^{(n_f)}\over Z_g^{(n_l)}}\zeta_g^0\right)^2\alpha_s^{(n_f)}(\mu)
\equiv\zeta_g^2\alpha_s^{(n_f)}(\mu),
\end{equation}
where $(Z_g)^2$ is the renormalization constant of the strong coupling constant.

The decoupling function plays an important role in testing QCD by running the
strong coupling to different energy scales.  For example, the strong coupling
$\alpha_s(m_{\tau})$ can be measured at the scale of the $\tau$-lepton mass
$m_{\tau}$. In the next step one can run this value to the scale of the
$Z$-boson mass $M_Z$ by using the proper running and decoupling at the heavy
charm- and bottom-quark thresholds and compare it to the experimentally
measured result of $\alpha_s(M_Z)$.  This procedure provides thus an excellent
test of QCD asymptotic freedom.  The four-loop contribution to the decoupling
function leads to a reduction of the matching-related uncertainties in the
evolution of $\alpha_s(m_{\tau})$ to $\alpha_s(M_Z)$ by a factor of
two~\cite{Chetyrkin:2005ia,Schroder:2005hy}.

\subsection*{Higgs-gluon coupling to N$^4$LO}

Gluon fusion is the dominant production mechanism of the SM Higgs boson $H$ at
the Large Hadron Collider (LHC), where the leading order process is already at
the one-loop level and the Higgs boson is produced by the fusion of two gluons
through a heavy top-quark loop.  The decoupling function enters in this context
as an important building block since it can be used to derive the effective
coupling of the Higgs boson to gluons via the following low-energy theorem
\begin{equation}
C_1^0=-{1\over2}m_t^0{\partial \ln\zeta_g^0 \over\partial m_t^0}
\,.
\end{equation}
$C_1$ enters into an effective Lagrangian
\begin{equation}
\mathcal{L}_{\mbox{\footnotesize eff}}=-2^{1/4}G_F^{1/2}HC_1[O_1']
\end{equation}
in QCD with five flavours, where the top mass dependence is contained in
$C_1$.  The symbol $G_F$ is the Fermi constant, and $[O_1']$ is the
renormalized form of the operator
$O_1'=G_{a\mu\nu}^{0'}G_{a}^{0'\mu\nu}$, where $G_{a\mu\nu}^{0'}$ is the
gluon field strength tensor. The prime indicates that the object is in
the effective five-flavour theory and the superscript $0$ denotes a bare
quantity. Using the four-loop result for $\zeta_g^2$ of
Refs.~\cite{Chetyrkin:2005ia,Schroder:2005hy} allows to determine $C_1$
in four-loop approximation, which confirms the result of
Ref.~\cite{Chetyrkin:1997un} in a completely different and independent
way. With the help of the anomalous dimensions even the five-loop
contribution to $C_1$ has been
predicted~\cite{Chetyrkin:2005ia,Schroder:2005hy} up to unknown
five-loop $n_f$-dependent terms of the $\beta$ function.

\subsection*{Decoupling of heavy quarks from the running of the fine
  structure constant}

\begin{figure}[!h]
\begin{center}
\begin{minipage}{2cm}
\begin{center}
\includegraphics[width=2cm]{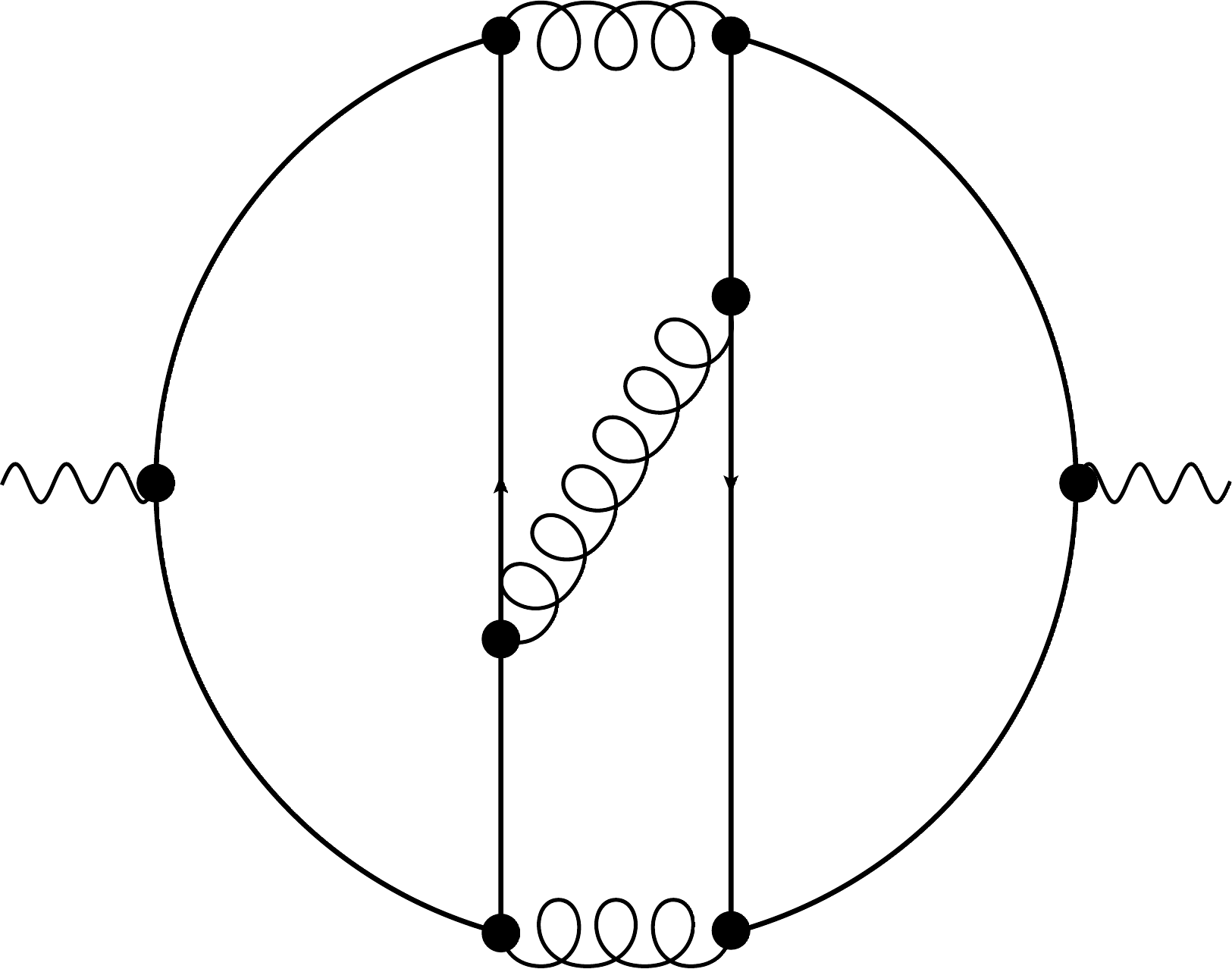}
\end{center}
\end{minipage}
\begin{minipage}{2cm}
\begin{center}
\includegraphics[width=2cm]{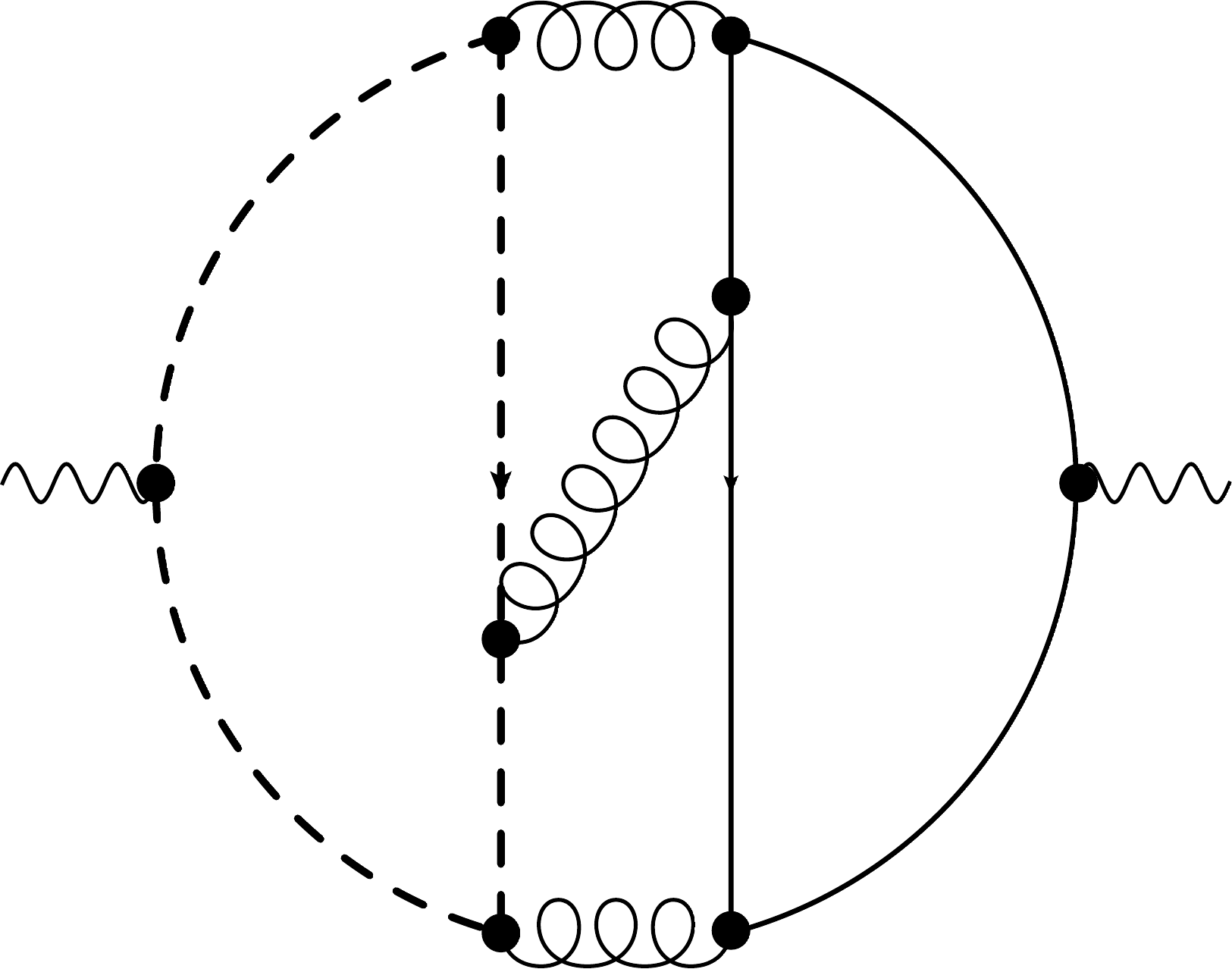}
\end{center}
\end{minipage}
\begin{minipage}{2cm}
\begin{center}
\includegraphics[width=2cm]{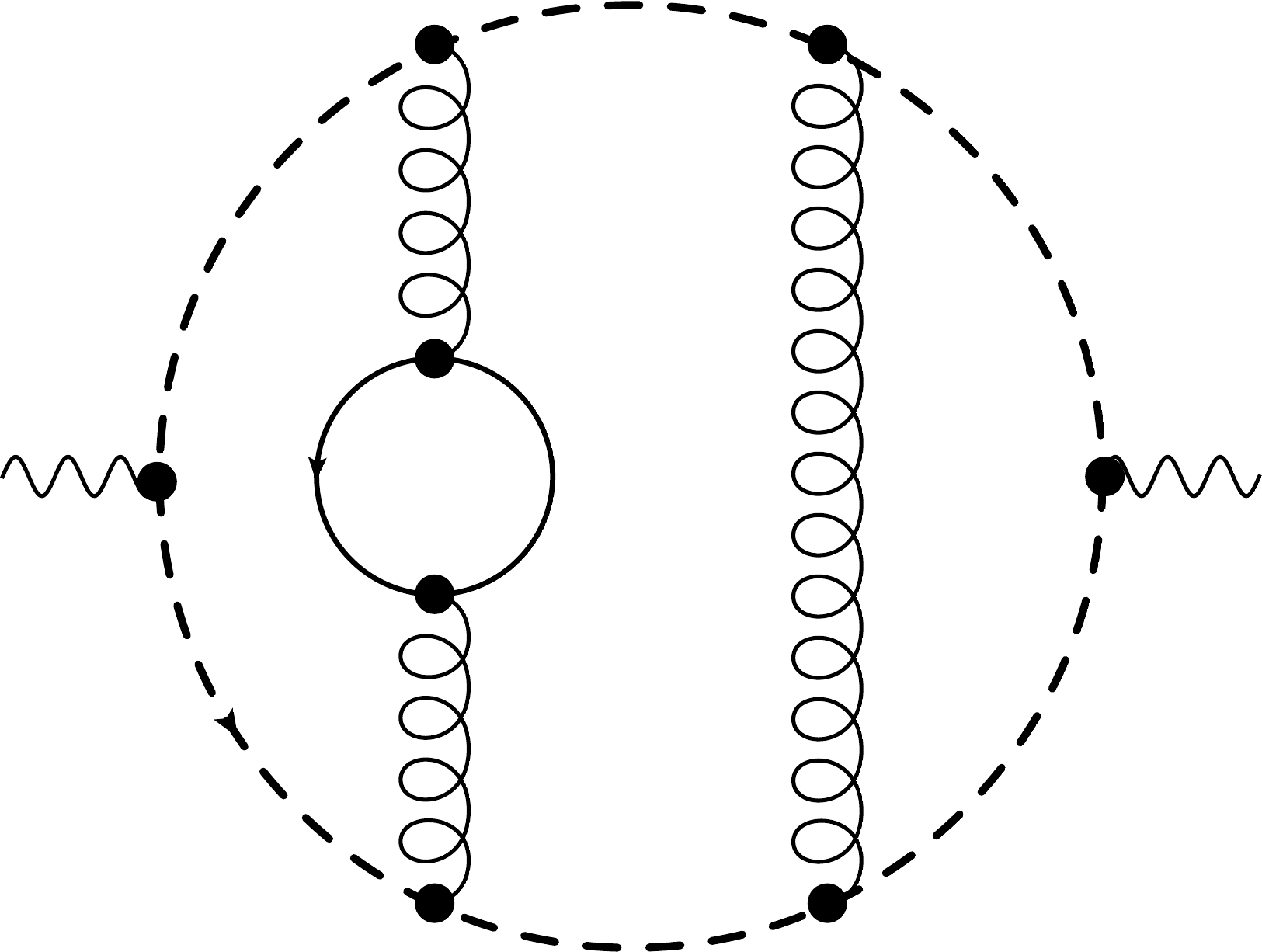}
\end{center}
\end{minipage}\\[0.2cm]
\begin{minipage}{2cm}
\begin{center}
\includegraphics[width=2cm]{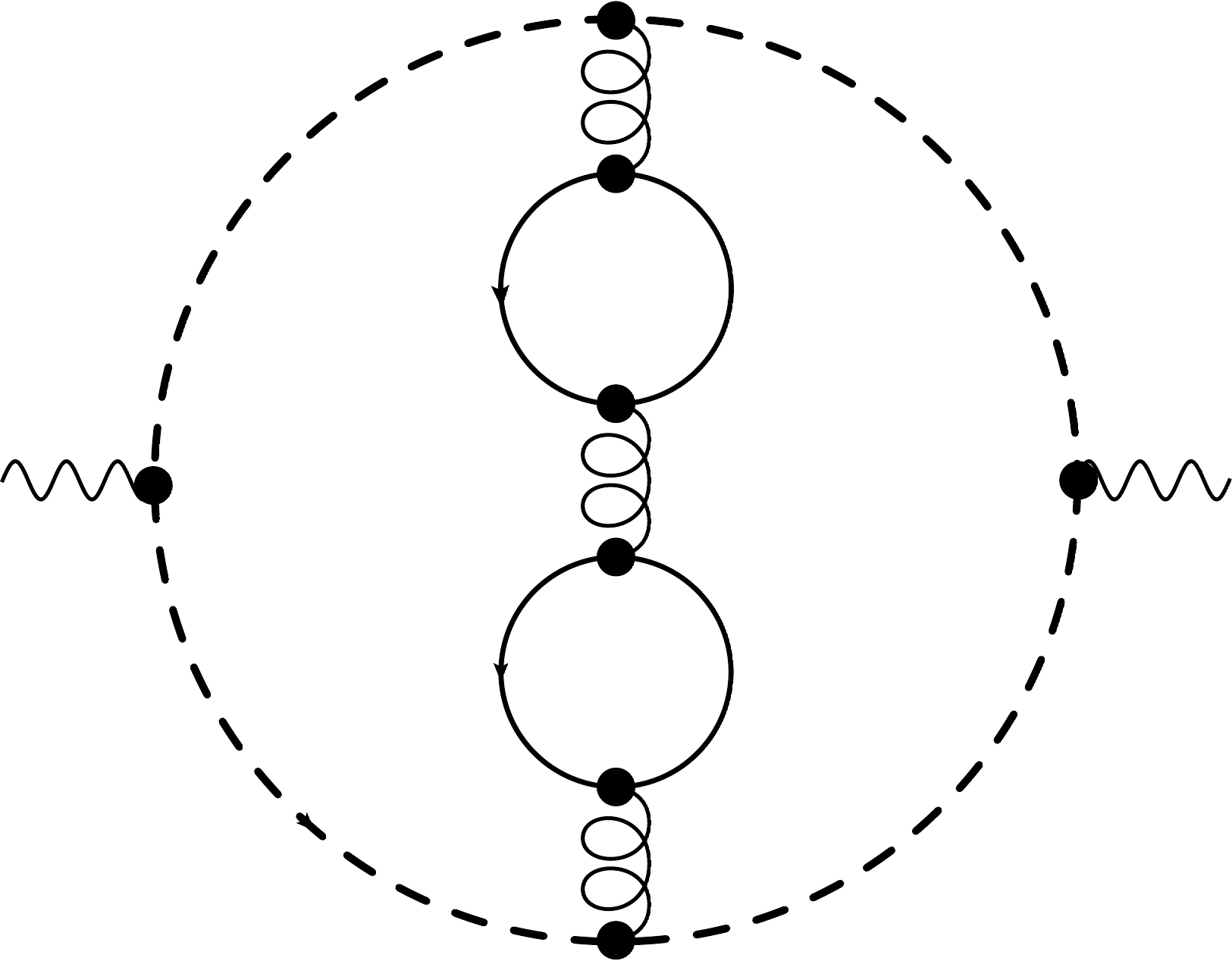}
\end{center}
\end{minipage}
\begin{minipage}{2cm}
\begin{center}
\includegraphics[width=2cm]{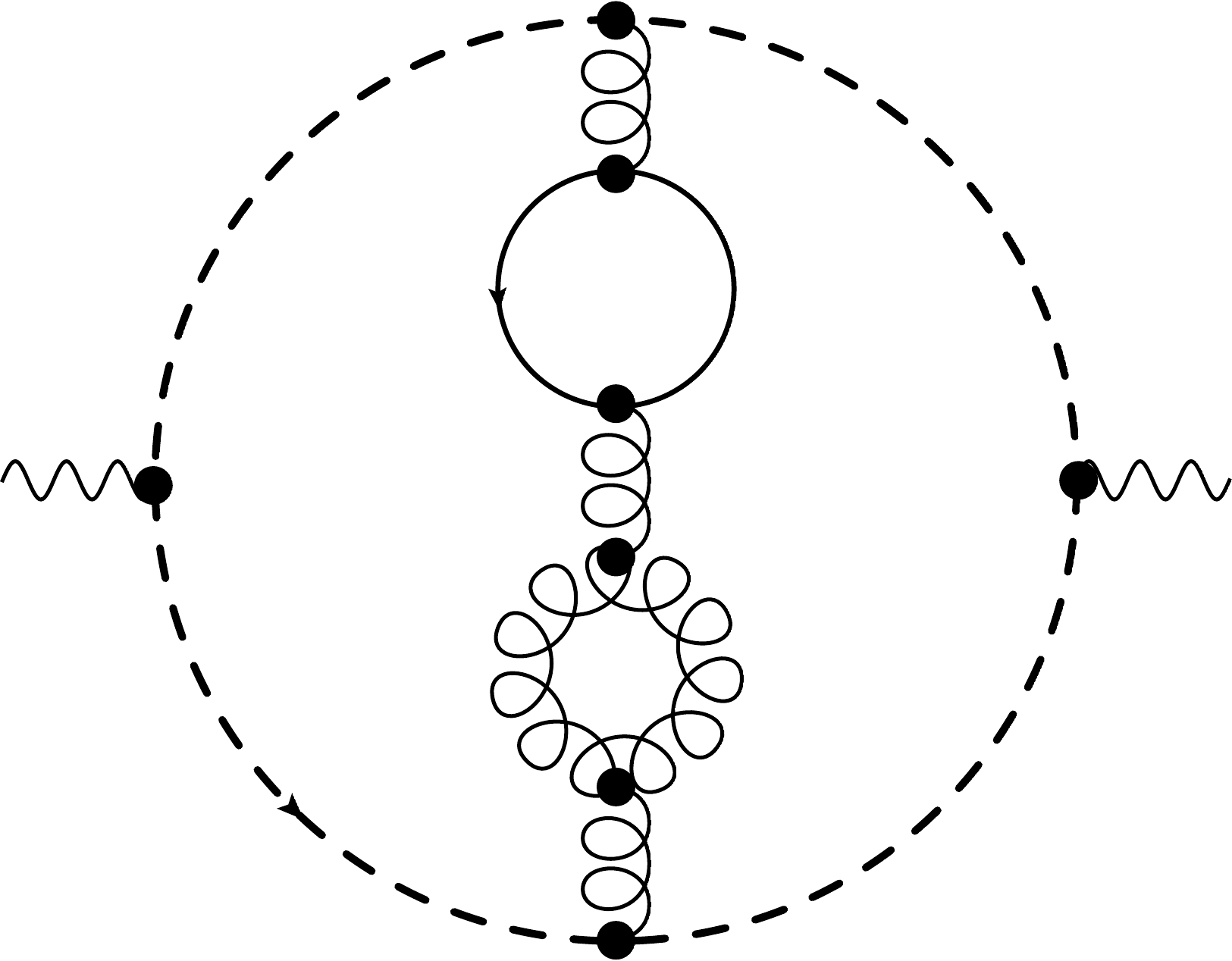}
\end{center}
\end{minipage}
\begin{minipage}{2cm}
\begin{center}
\includegraphics[width=2cm]{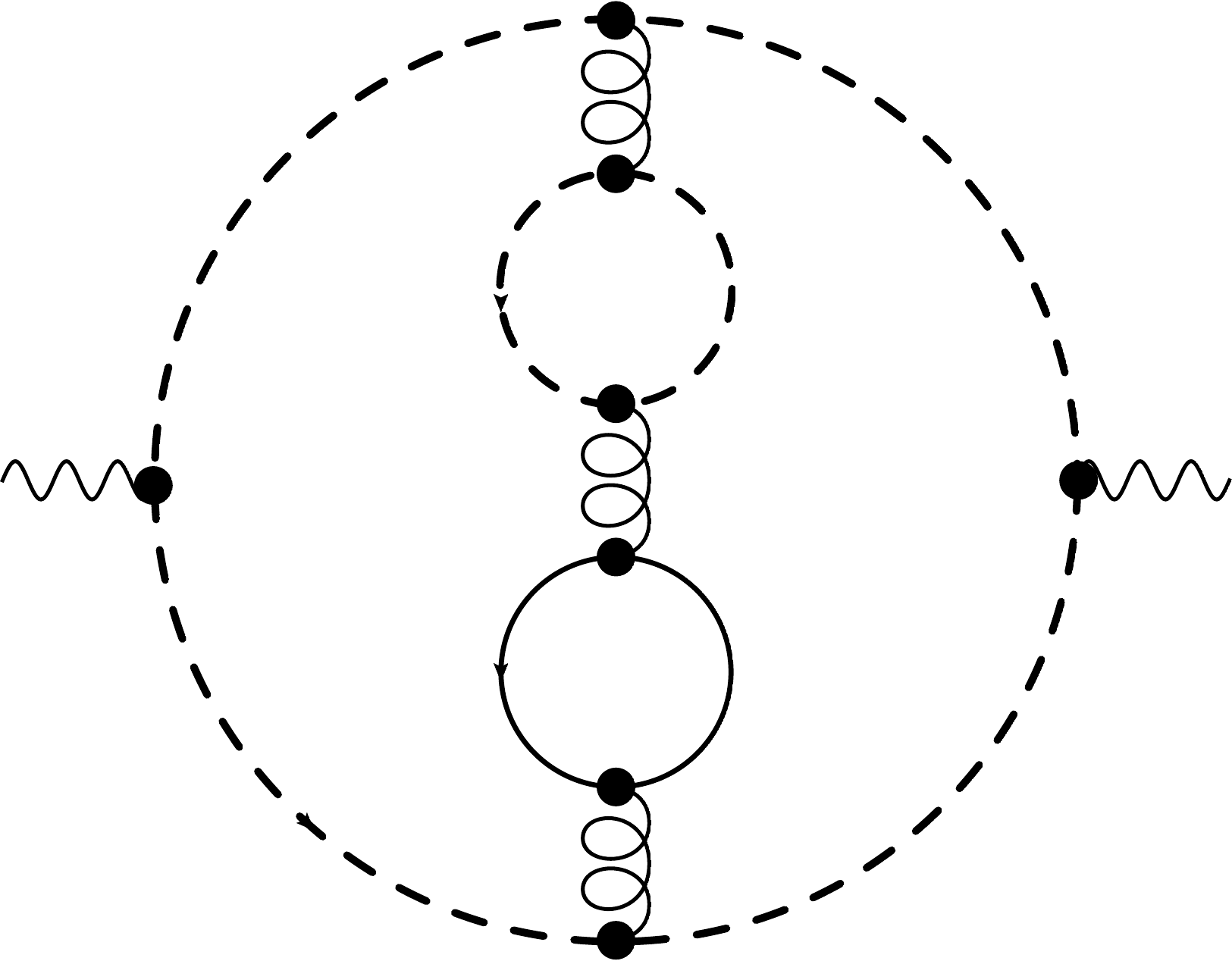}
\end{center}
\end{minipage}
\end{center}
\vspace*{-0.6cm}
\caption{
The first two diagrams are example singlet diagrams; the last four
diagrams are examples for the situation where the external photon
couples to a massless fermion loop with the insertion of a heavy
internal fermion loop.  The solid lines represent heavy top-quarks, the
twisted lines denote gluons, and the dashed lines represent massless
quarks.\label{fig:Pi4loop}}
\end{figure}

In complete analogy one can determine from the massive photon vacuum
polarization function the photon decoupling function
${\left({\zeta^0_{g\gamma}}\right)^2}$ 
\begin{equation}
{\left({\zeta^0_{g\gamma}}\right)^2}={1\over 1+\Pi_{\gamma}^{0h}(0)}\,.
\end{equation}
The three-loop results of Ref.~\cite{Chetyrkin:1997un} have been
extended to four loops in Ref.~\cite{Sturm:2014nva}. Starting from
three-loop order there arise also diagrams where the external photon couples
to massless fermions with the insertion of a heavy fermion loop. At
four-loop order also singlet type diagrams arise for the first time,
where the photon couples to two different fermion loops. Some example
diagrams are shown in Fig.~\ref{fig:Pi4loop}.
%
%

\subsection*{Higgs boson decay to photons}

The photon decoupling function can be used to determine higher order
QCD corrections to the partial decay width of the Higgs boson into two
photons ($\gamma$). The amplitude of the partial decay width $H\to\gamma\gamma$

\begin{equation}
\label{eq:decaywidth}
\Gamma(H\to\gamma\gamma)={M_H^3\over64\*\pi}\*\Big|A_W(\tau_W)+\sum_{f}A_f(\tau_f)\Big|^2\,,
\end{equation}
consists of two parts, a purely bosonic part $A_W(\tau_W)$ and a
purely fermionic part $A_f(\tau_f)$ with $\tau_W=M_H^2/(4M_W^2)$ and
$\tau_f=M_H^2/(4M_f^2)$. Higher order QCD corrections modify the
fermionic part $A_f(\tau_f)$ of the amplitude. The top quark gives a
dominant contribution to the amplitude $A_f$ ($f=t$), since it is the heaviest
fermion in the SM. In the heavy top-quark mass limit one can again
describe the Higgs-photon-photon interaction in terms of an effective
Lagrangian approach 
\begin{equation}
\label{eq:Lhgg}
\mathcal{L}_{\mbox{\scriptsize{eff}}}=-{H^0\over v^0}\,C^0_{1\gamma}\,F^{'0,\mu\nu}F^{'0}_{\mu\nu}\,,
\end{equation}
with the vacuum expectation value $v^0$ and the field strength tensor
$F_{\mu\nu}^{'0}$. The subscript $0$ indicates a bare quantity and the
prime denotes that the quantity is considered in the effective theory
with $n_l$ light active quark flavours. The coefficient function
$C_{1\gamma}^{0}$ depends on the photon decoupling function
\begin{equation}
C_{1\gamma}^{0}=-{1\over2}m_t^0{\partial\ln \zeta_{g\gamma}^0 \over\partial
m_t^0}.
\end{equation}
This approach allows one to determine the leading contributions in the
heavy top-quark mass limit to the Higgs-boson decay into two photons,
where the external photons couple to the same heavy fermion loop, the so
called non-singlet contributions.

\begin{figure}[!h]
\begin{center}
\includegraphics[width=4cm]{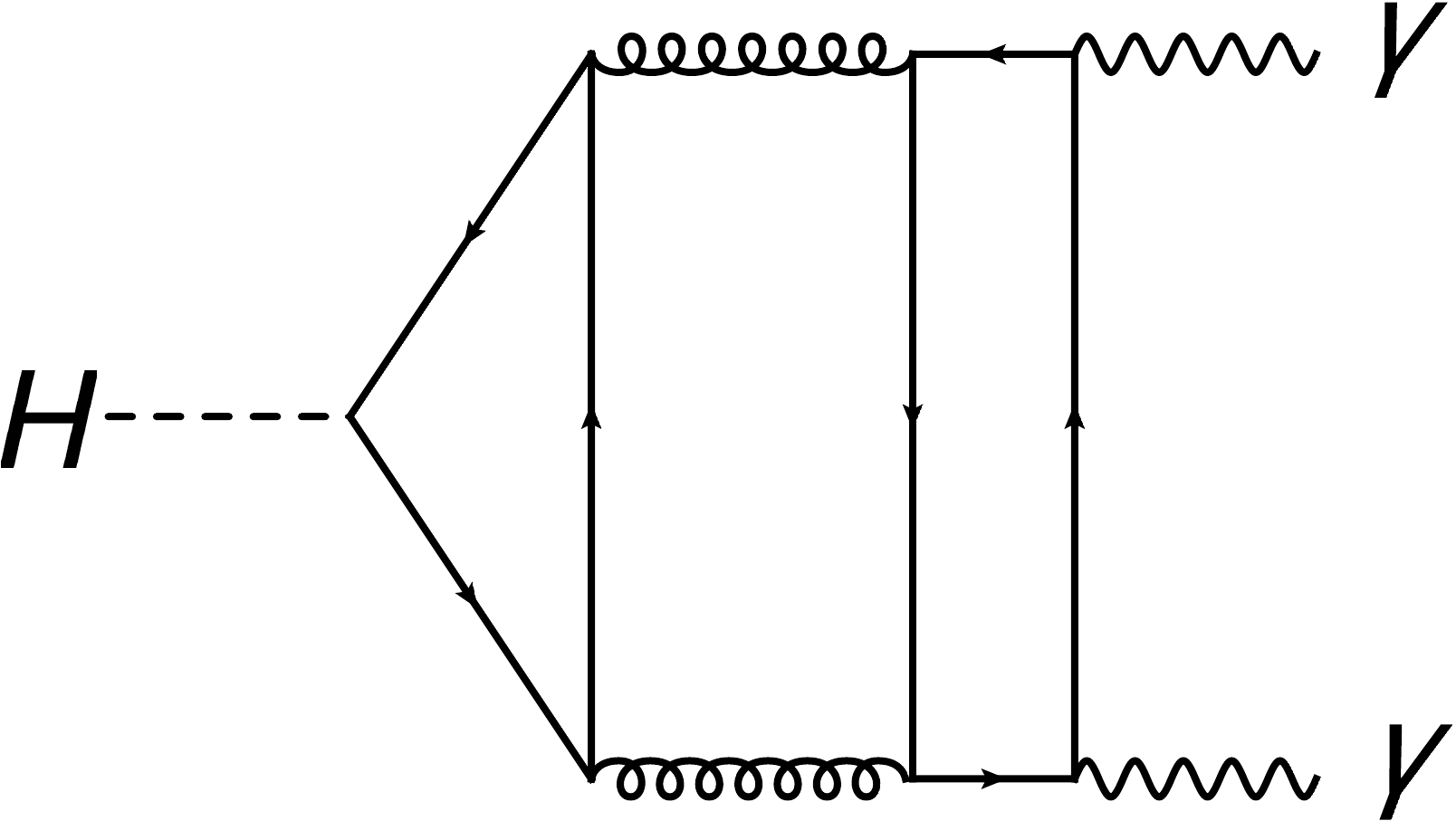}
\end{center}
\vspace*{-0.6cm}
\caption{Example three-loop singlet diagram. Solid lines denote top
  quarks, wavy lines are photons, twisted lines represent gluons and the
  dashed line is the Higgs boson.\label{fig:Hgamgam3loopSing} }
\end{figure}

At three-loop order in perturbative QCD the non-singlet contributions to the
decay $H\to\gamma\gamma$ have been computed in
Ref.~\cite{Steinhauser:1996wy} with several different methods, including
power corrections of the order $[M_H^2/(4M_t^2)]^2$. The singlet
contributions have been added  in Ref.~\cite{Maierhofer:2012vv},
where also additional power corrections of higher orders in
$M_H^2/(4M_t^2)$ were calculated.  The singlet contributions appear for
the first time at three-loop order. An example diagram is shown in
Fig.~\ref{fig:Hgamgam3loopSing}.

\begin{figure}[!h]
\begin{center}
\begin{minipage}{2.5cm}
\begin{center}
\includegraphics[width=2.5cm]{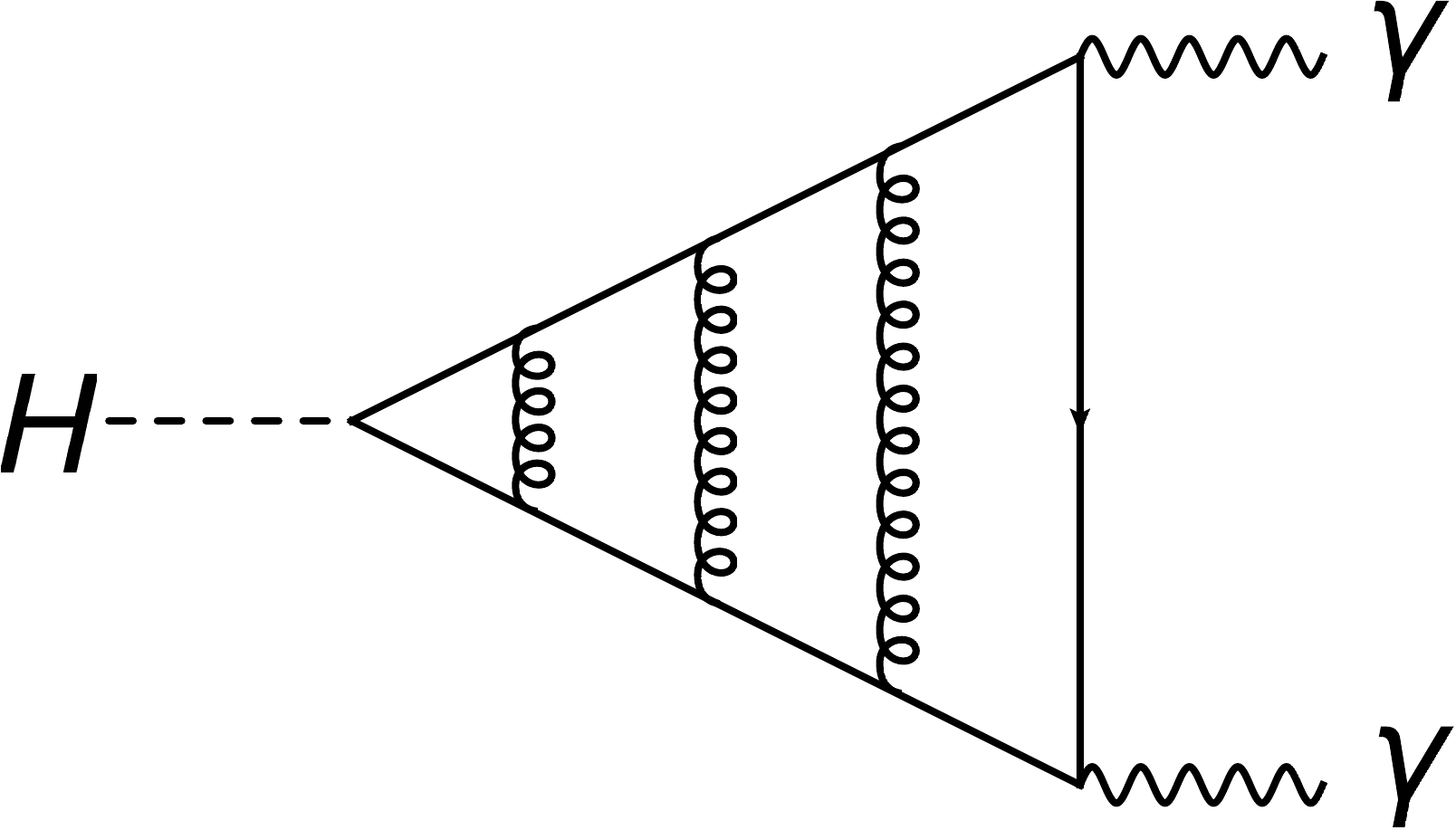}\\
\end{center}
\end{minipage}
\begin{minipage}{2.5cm}
\begin{center}
\includegraphics[width=2.5cm]{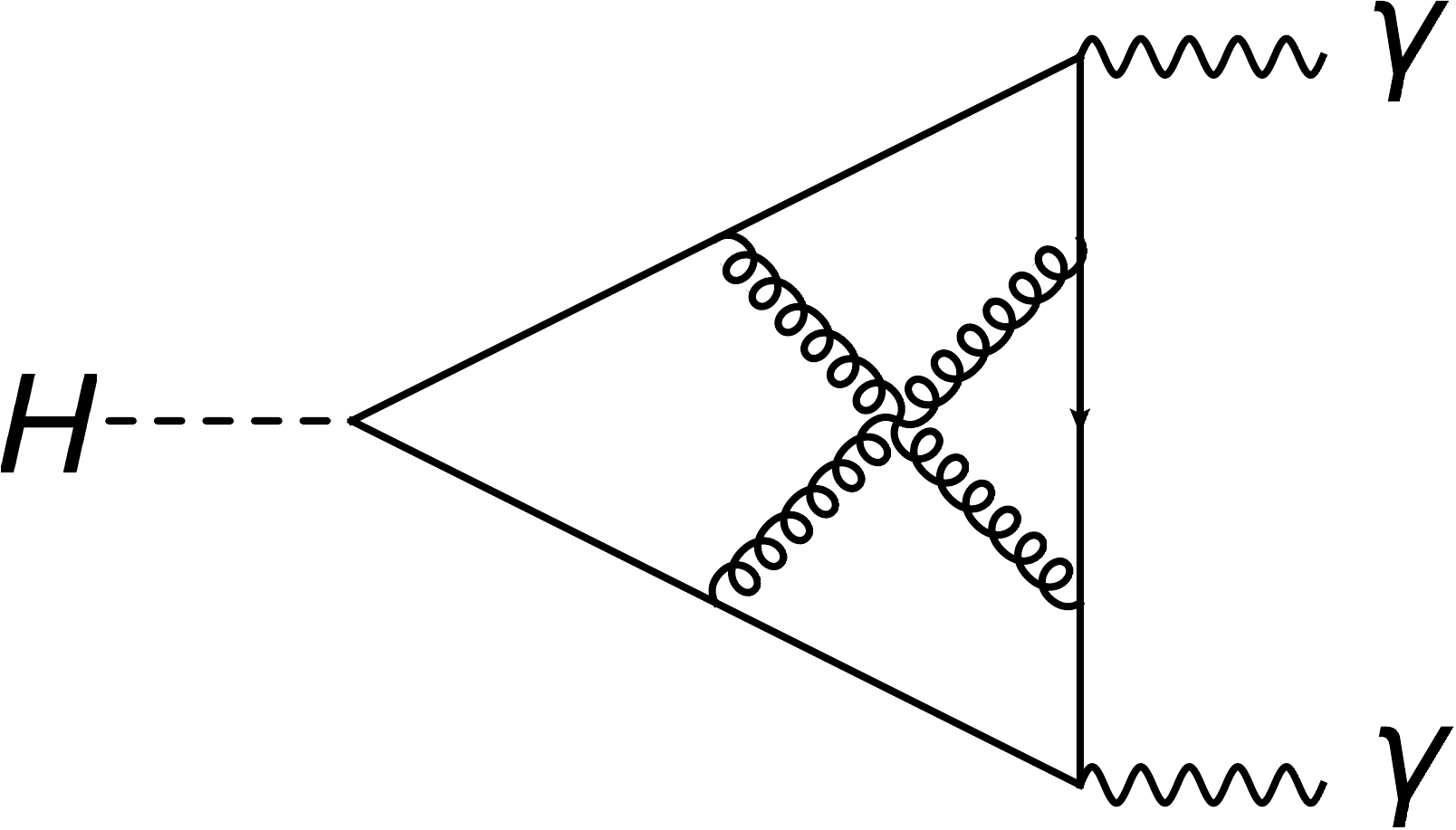}\\
\end{center}
\end{minipage}
\begin{minipage}{2.5cm}
\begin{center}
\includegraphics[width=2.5cm]{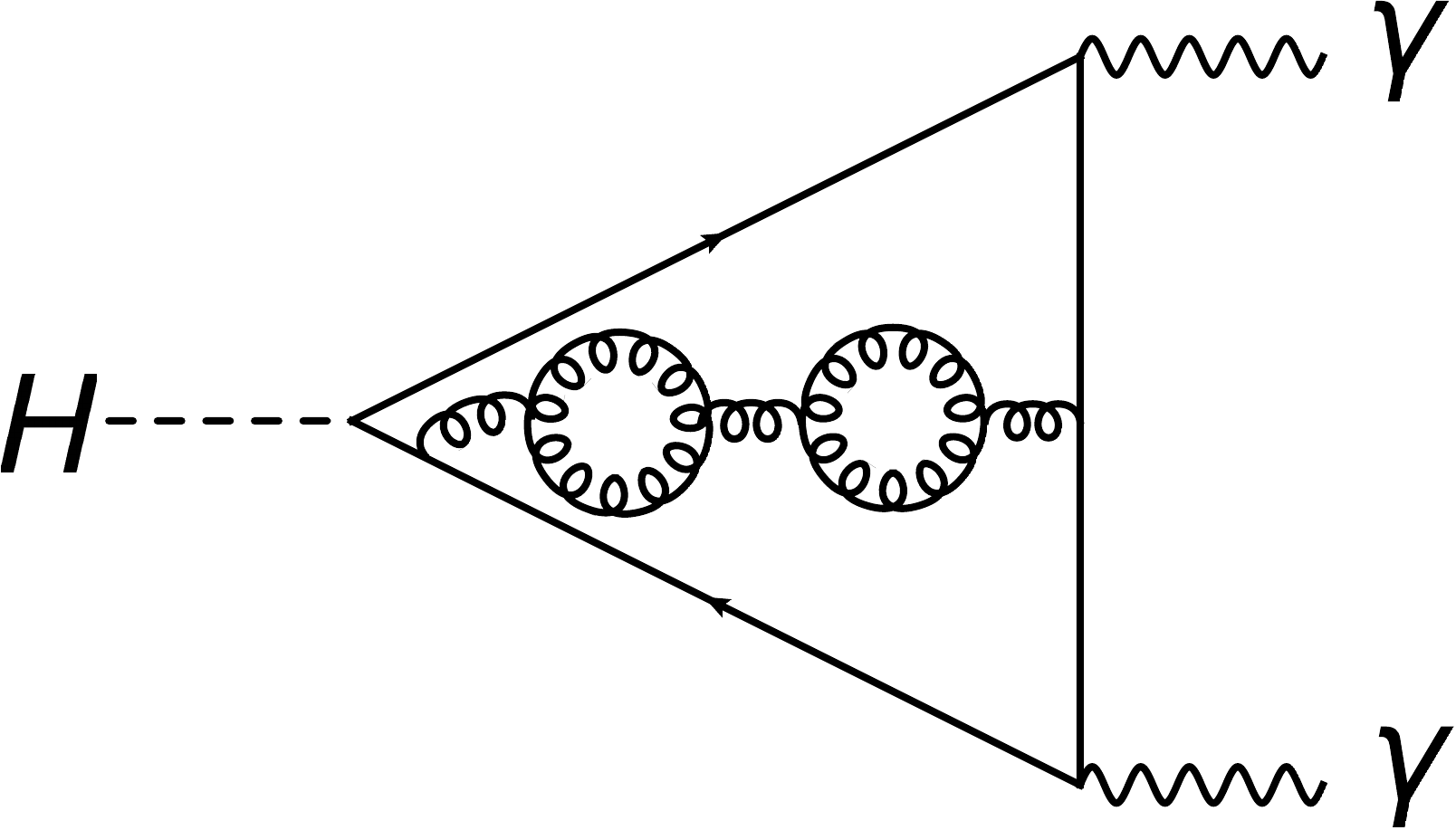}\\
\end{center}
\end{minipage}\\[0.2cm]
%
%
%
%
\begin{minipage}{2.5cm}
\begin{center}
\includegraphics[width=2.5cm]{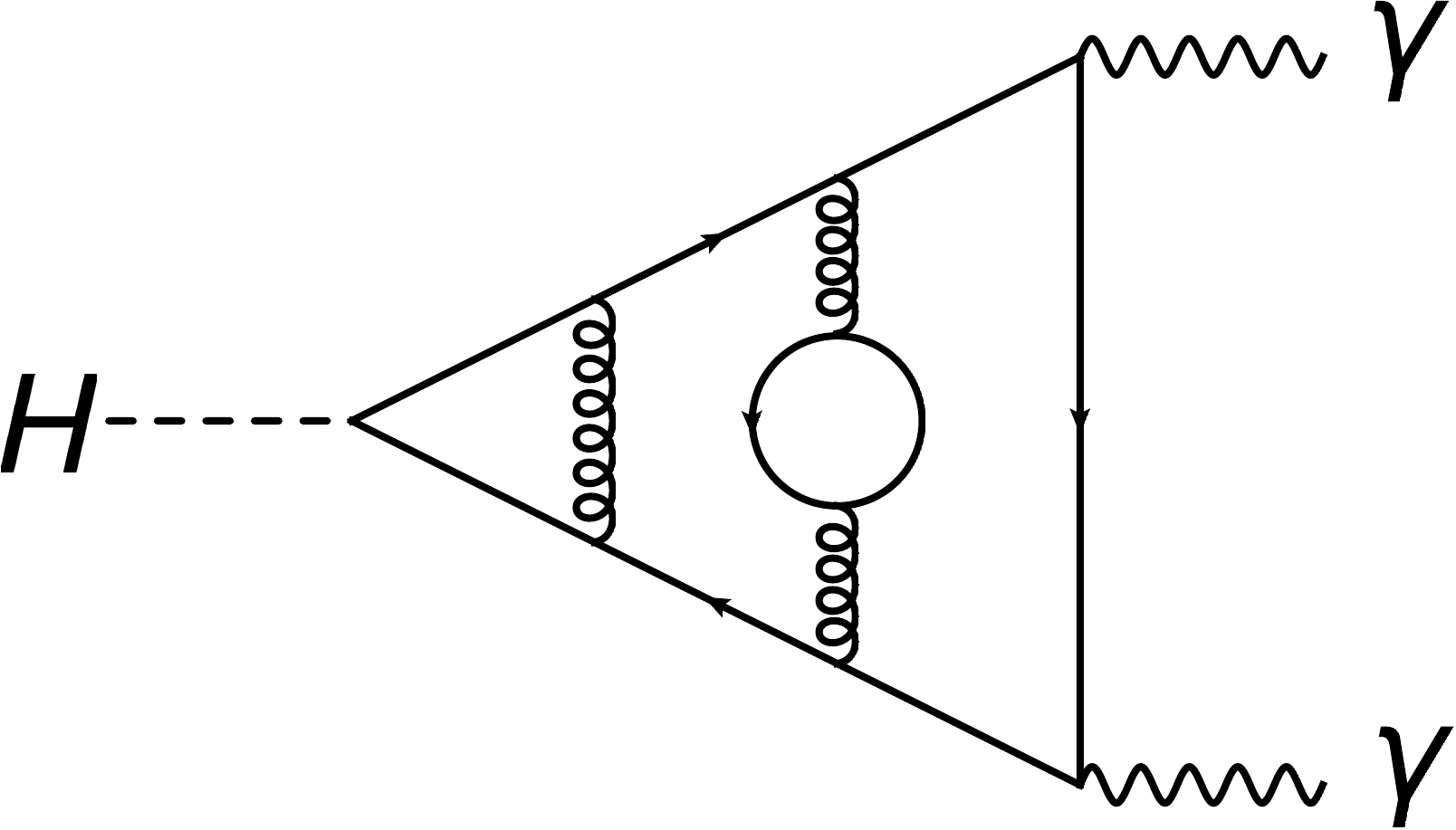}\\
\end{center}
\end{minipage}
\begin{minipage}{2.5cm}
\begin{center}
\includegraphics[width=2.5cm]{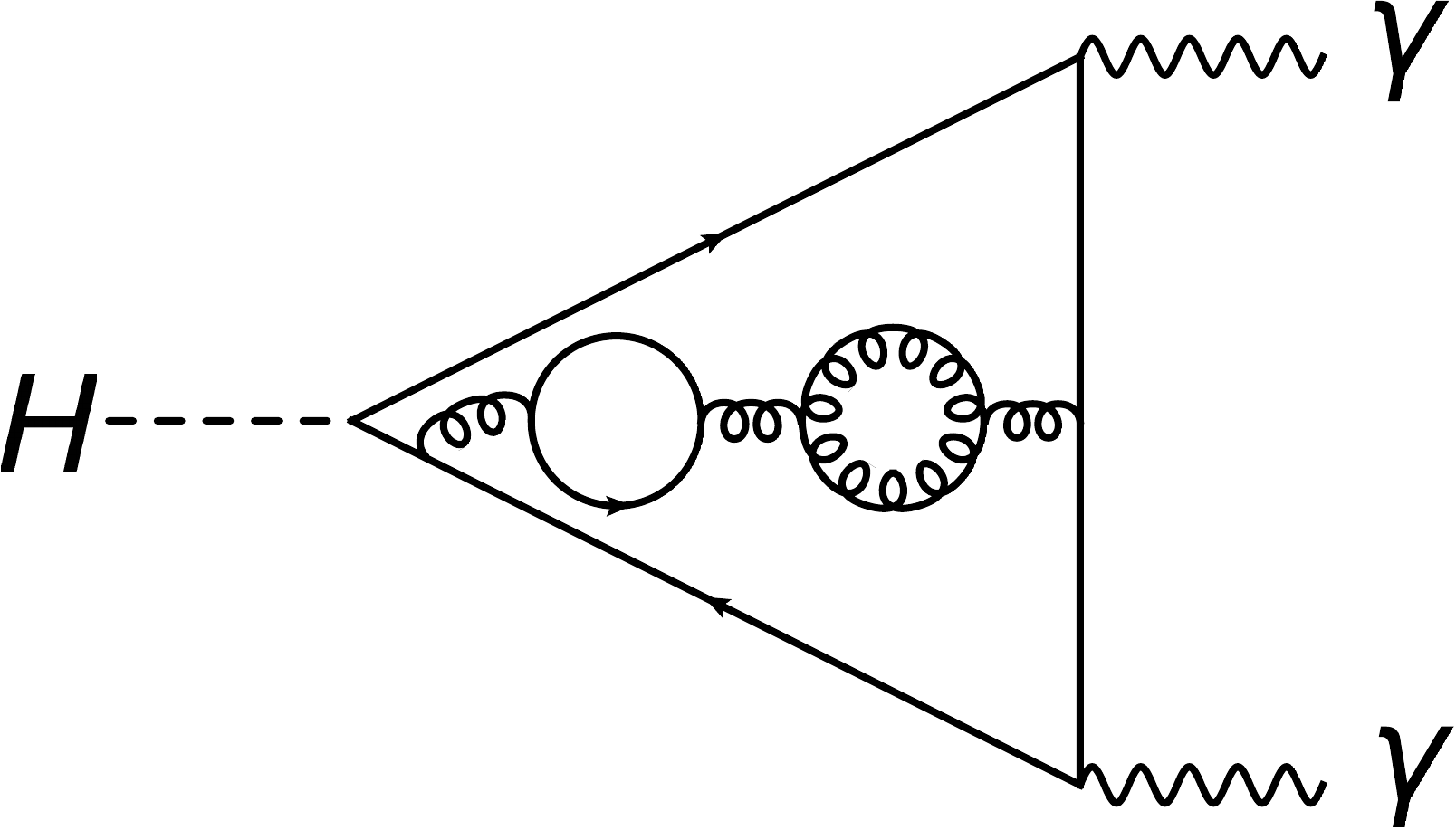}\\
\end{center}
\end{minipage}
\begin{minipage}{2.5cm}
\begin{center}
\includegraphics[width=2.5cm]{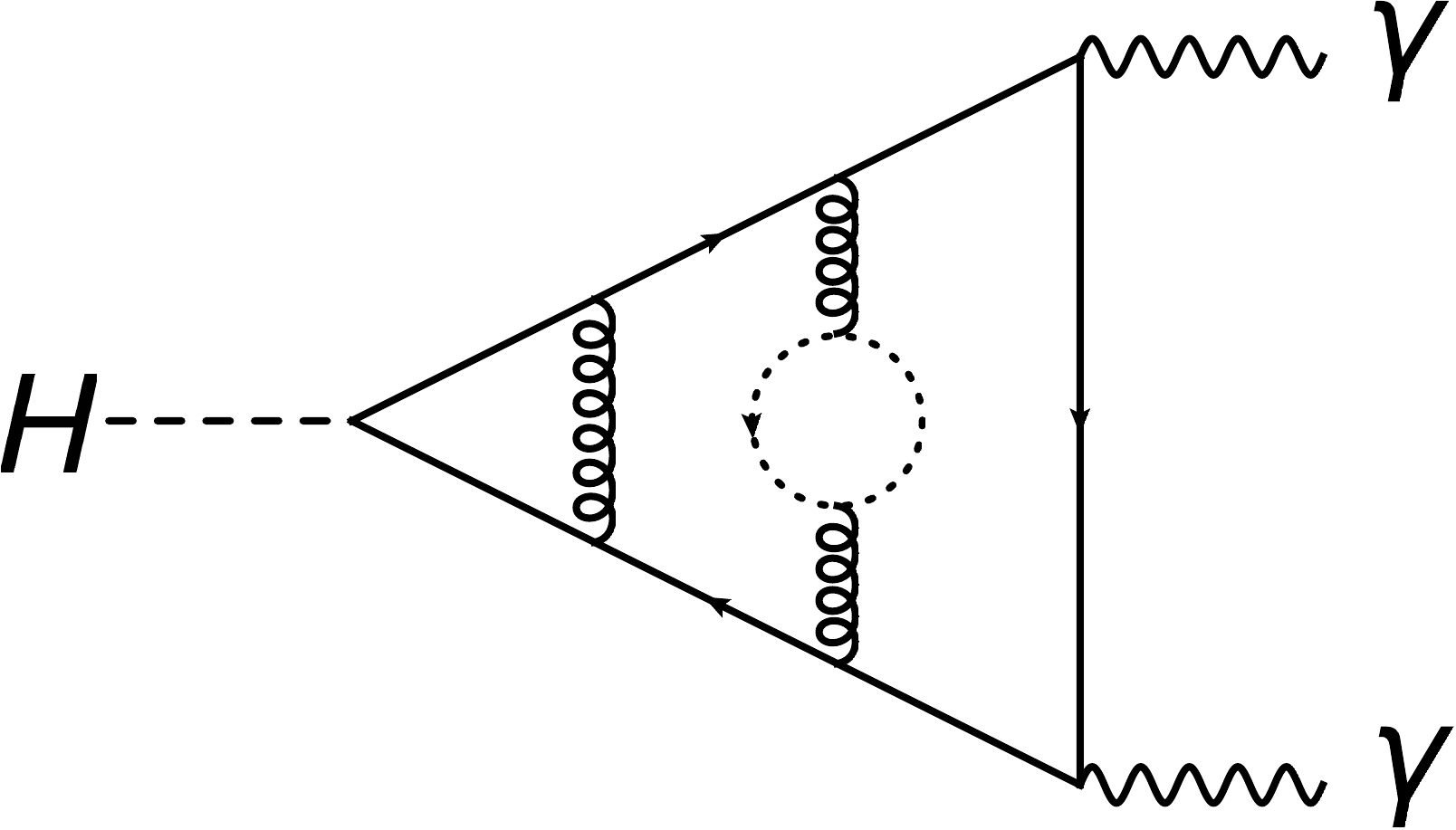}\\
\end{center}
\end{minipage}\\[0.2cm]
%
%
%
%
\begin{minipage}{2.5cm}
\begin{center}
\includegraphics[width=2.5cm]{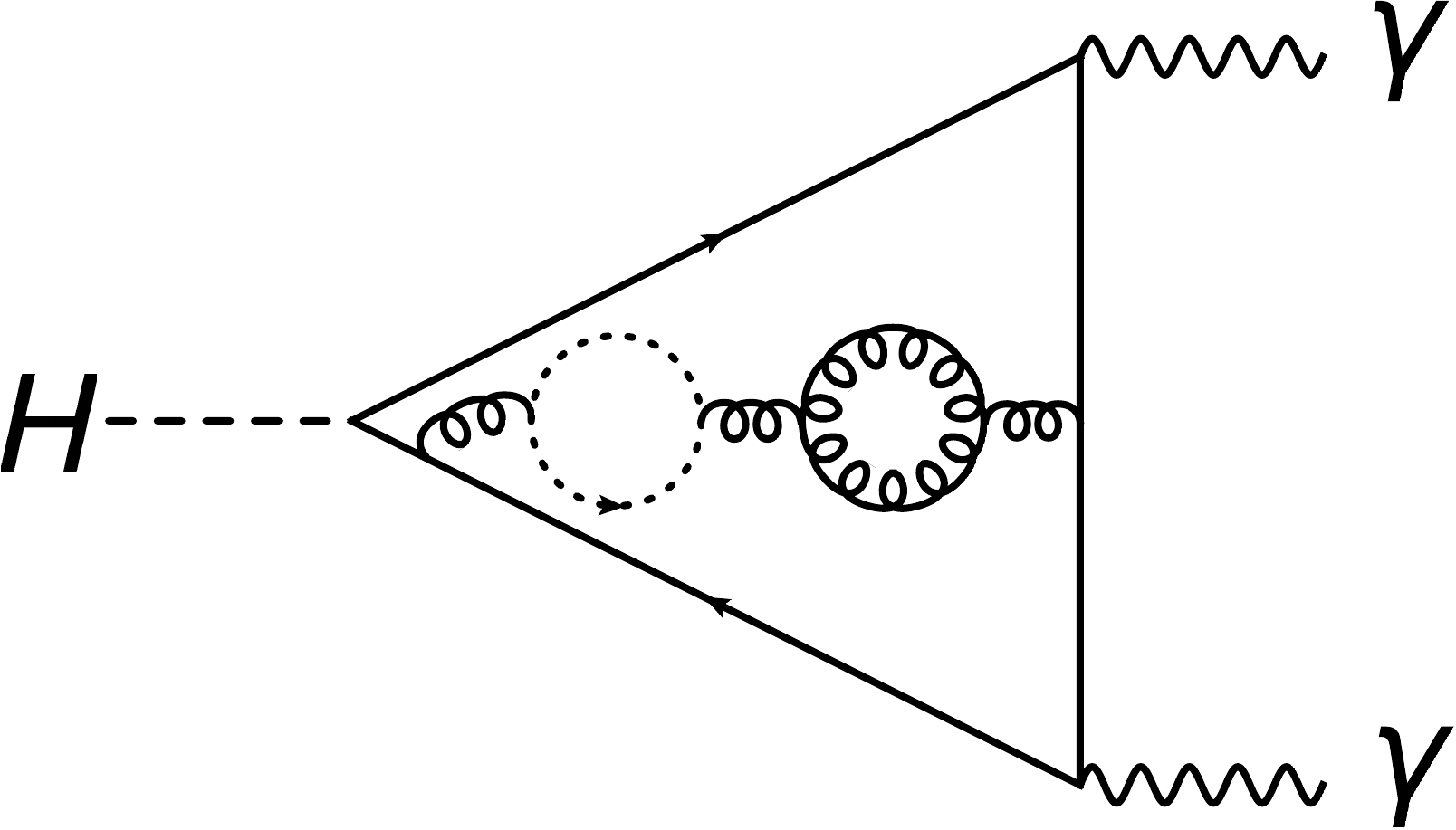}\\
\end{center}
\end{minipage}
\begin{minipage}{2.5cm}
\begin{center}
\includegraphics[width=2.5cm]{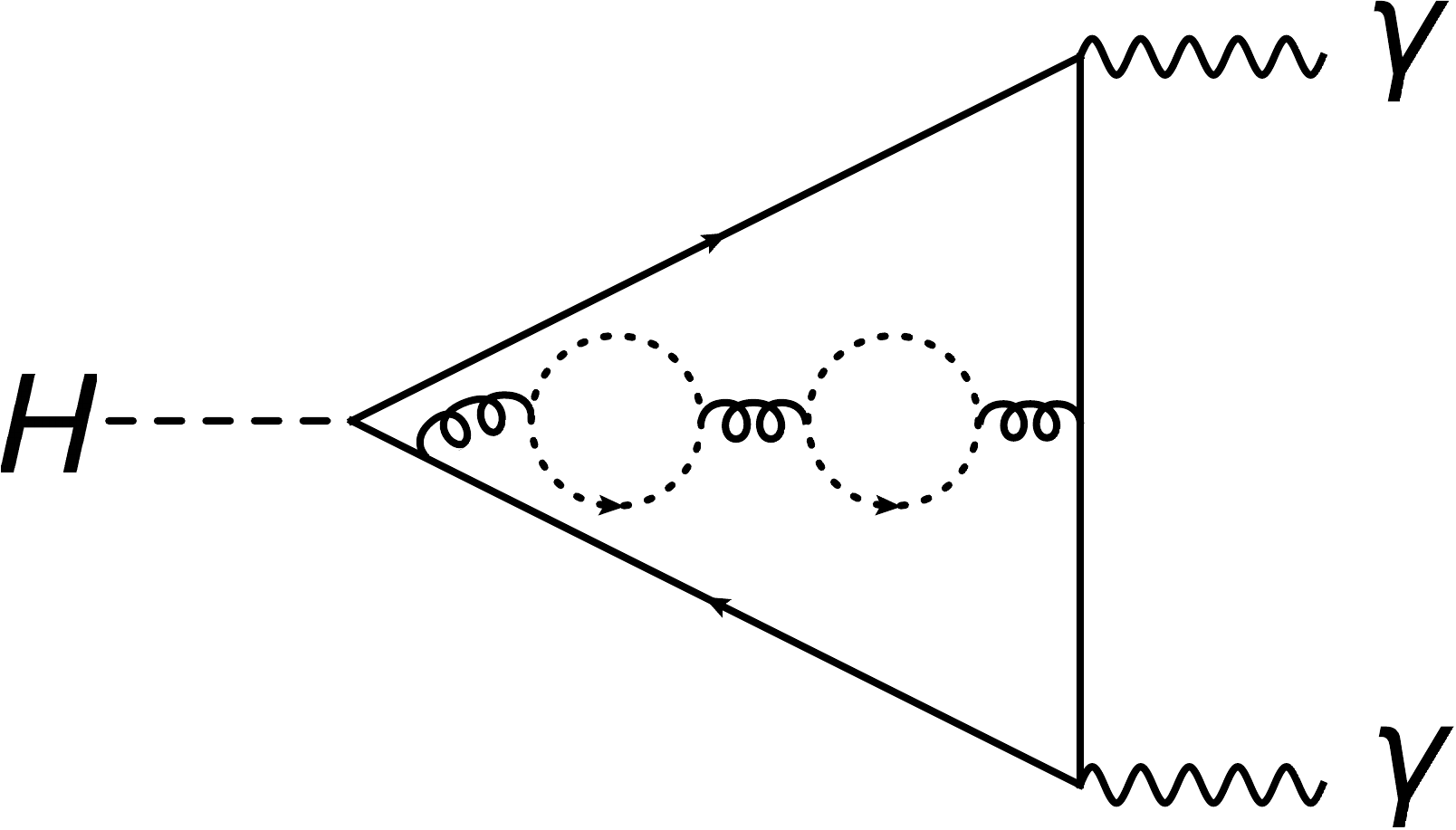}\\
\end{center}
\end{minipage}
\begin{minipage}{2.5cm}
\begin{center}
\includegraphics[width=2.5cm]{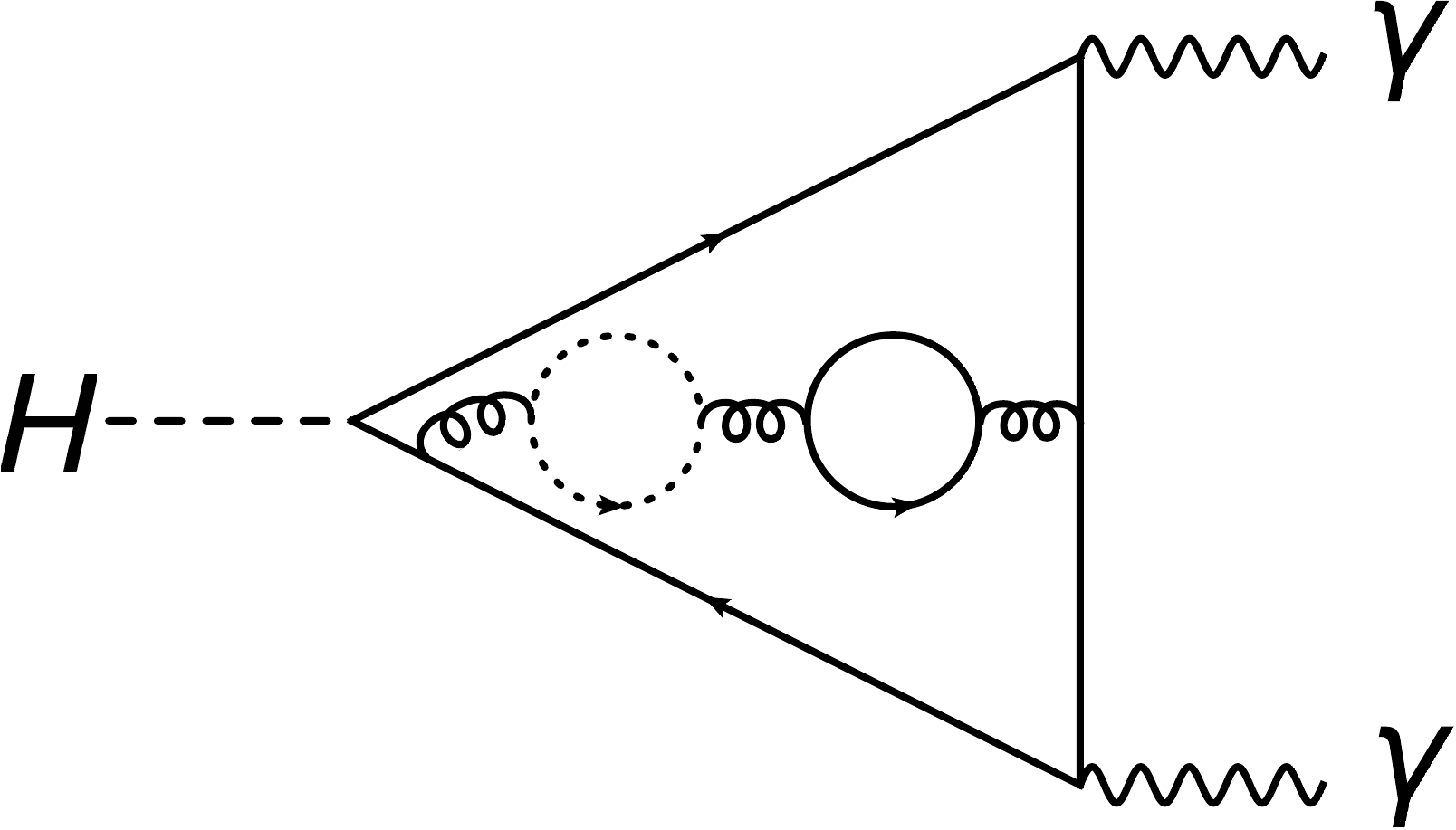}\\
\end{center}
\end{minipage}\\[0.2cm]
%
%
%
%
\begin{minipage}{2.5cm}
\begin{center}
\includegraphics[width=2.5cm]{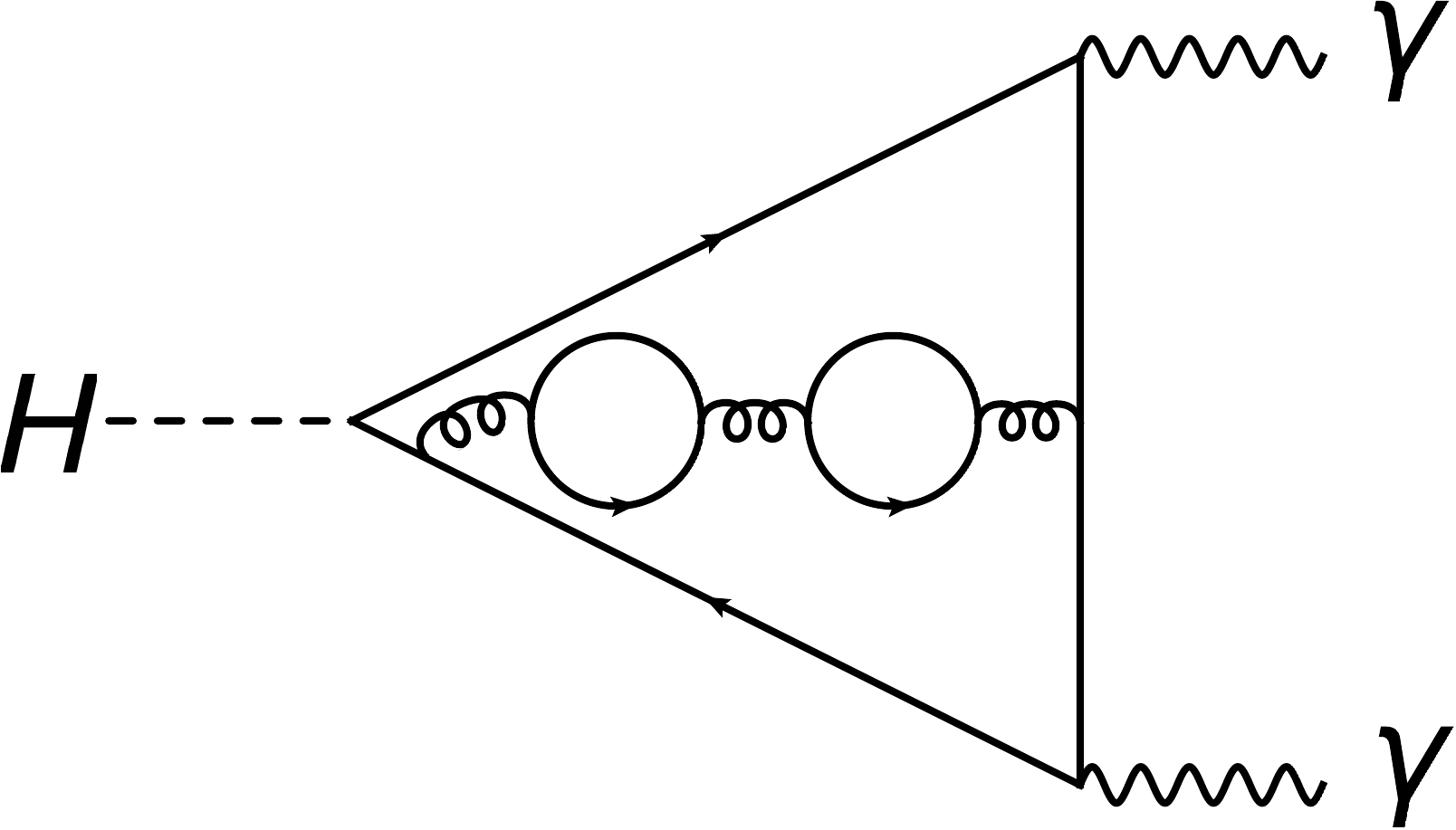}\\
\end{center}
\end{minipage}
\begin{minipage}{2.5cm}
\begin{center}
\includegraphics[width=2.5cm]{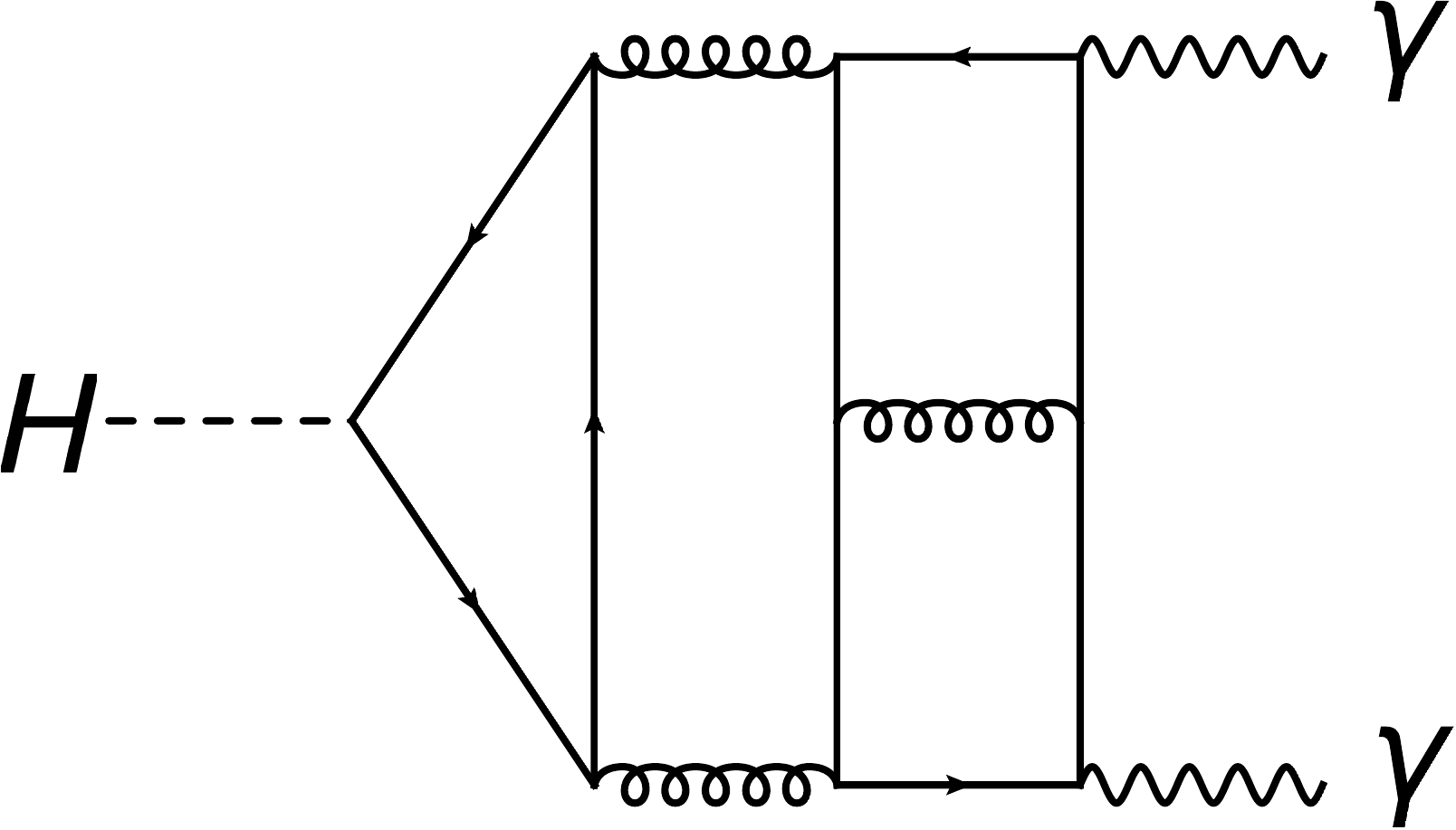}\\
\end{center}
\end{minipage}
\begin{minipage}{2.5cm}
\begin{center}
\includegraphics[width=2.5cm]{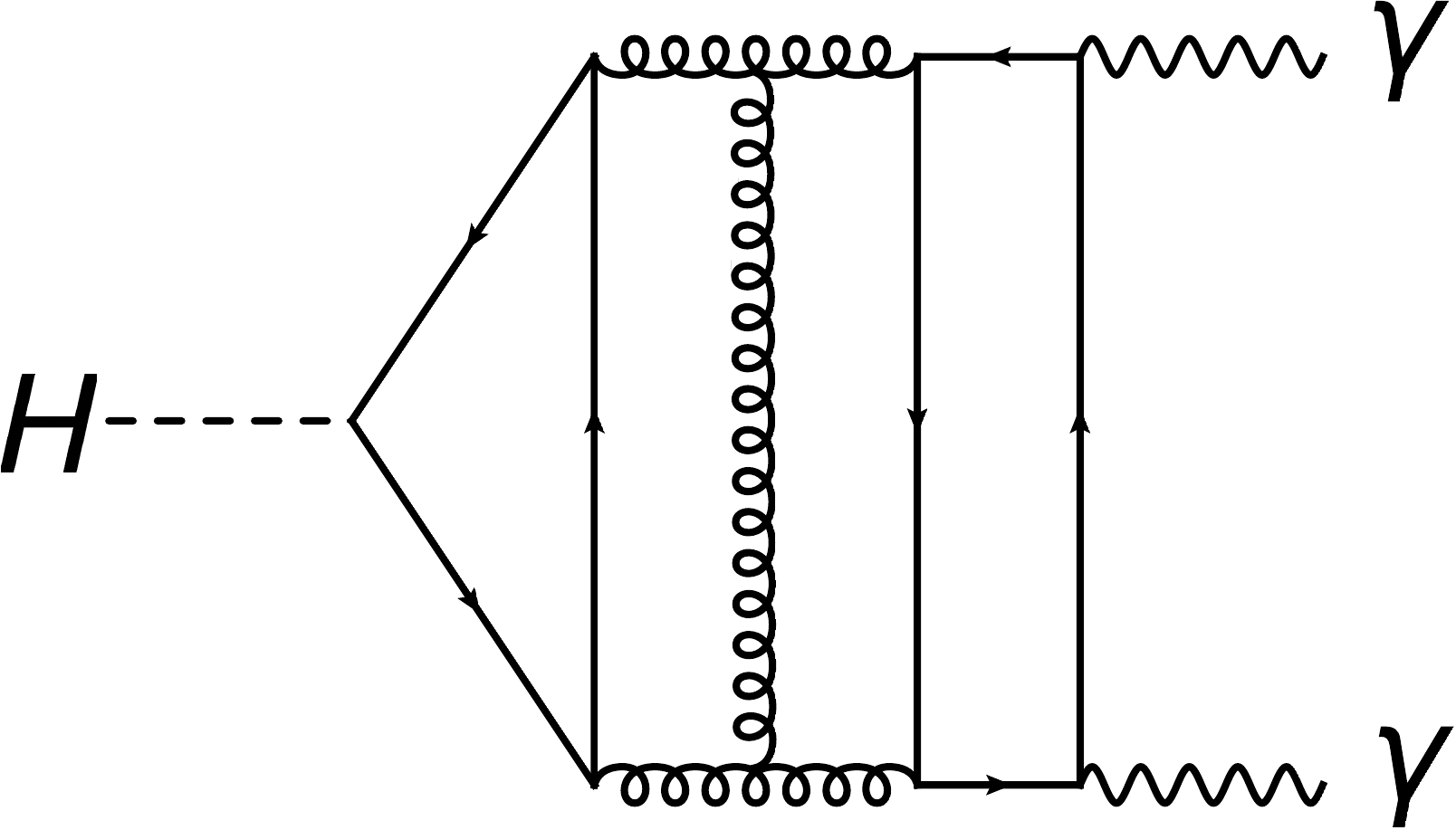}\\
\end{center}
\end{minipage}\\[0.2cm]
%
%
%
%
\begin{minipage}{2.5cm}
\begin{center}
\includegraphics[width=2.5cm]{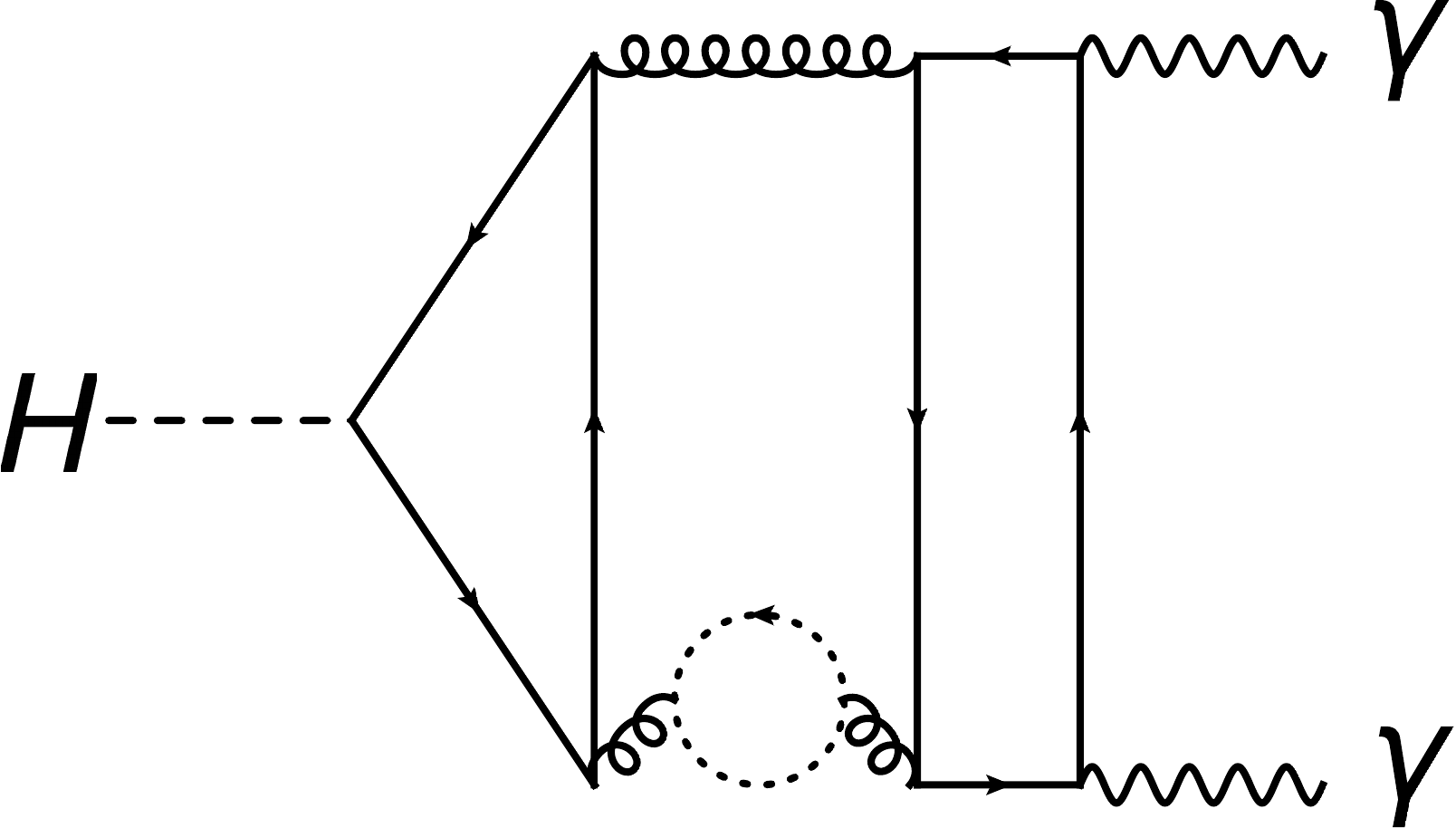}\\
\end{center}
\end{minipage}
\begin{minipage}{2.5cm}
\begin{center}
\includegraphics[width=2.5cm]{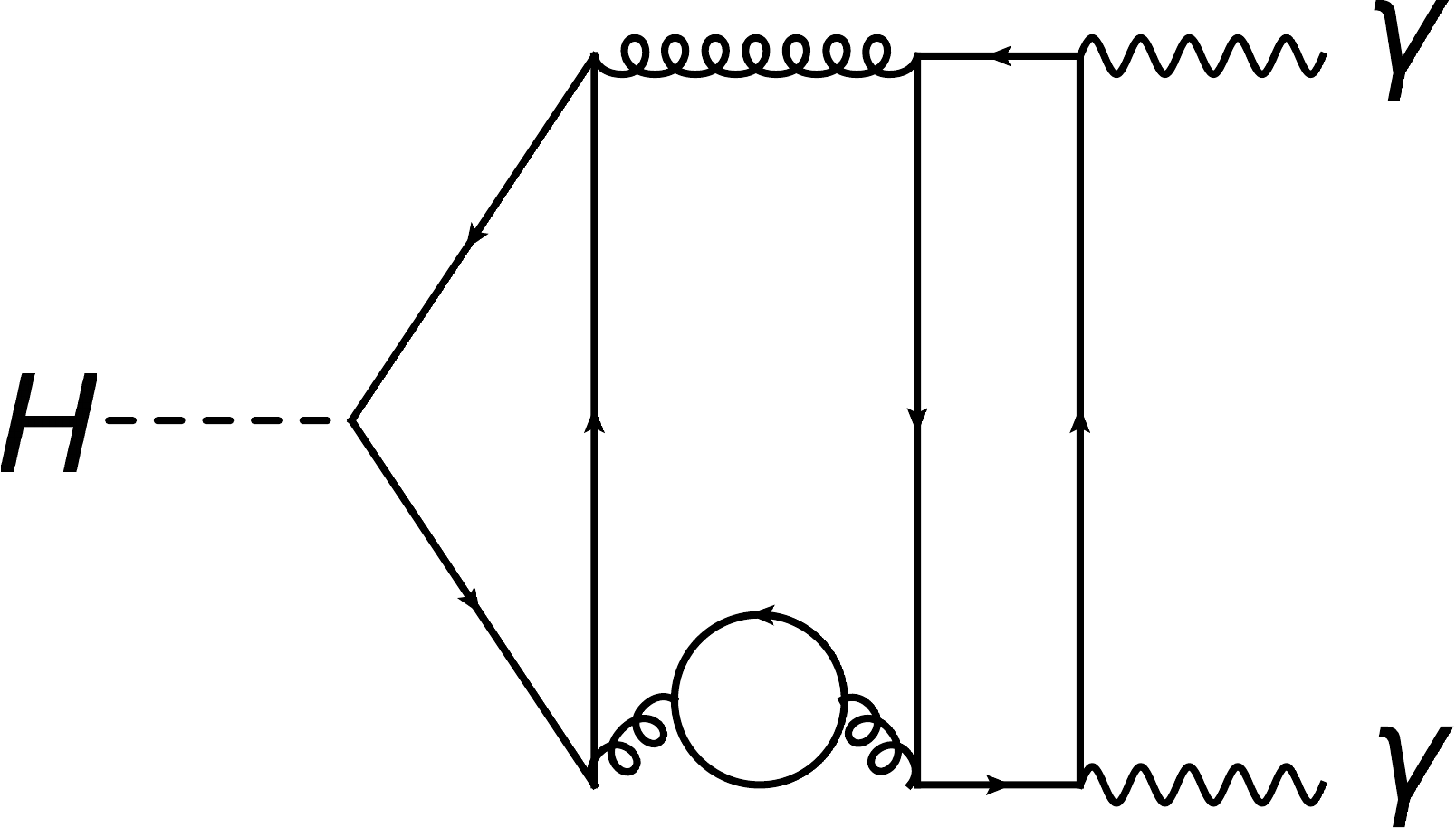}\\
\end{center}
\end{minipage}
\begin{minipage}{2.5cm}
\begin{center}
\includegraphics[width=2.5cm]{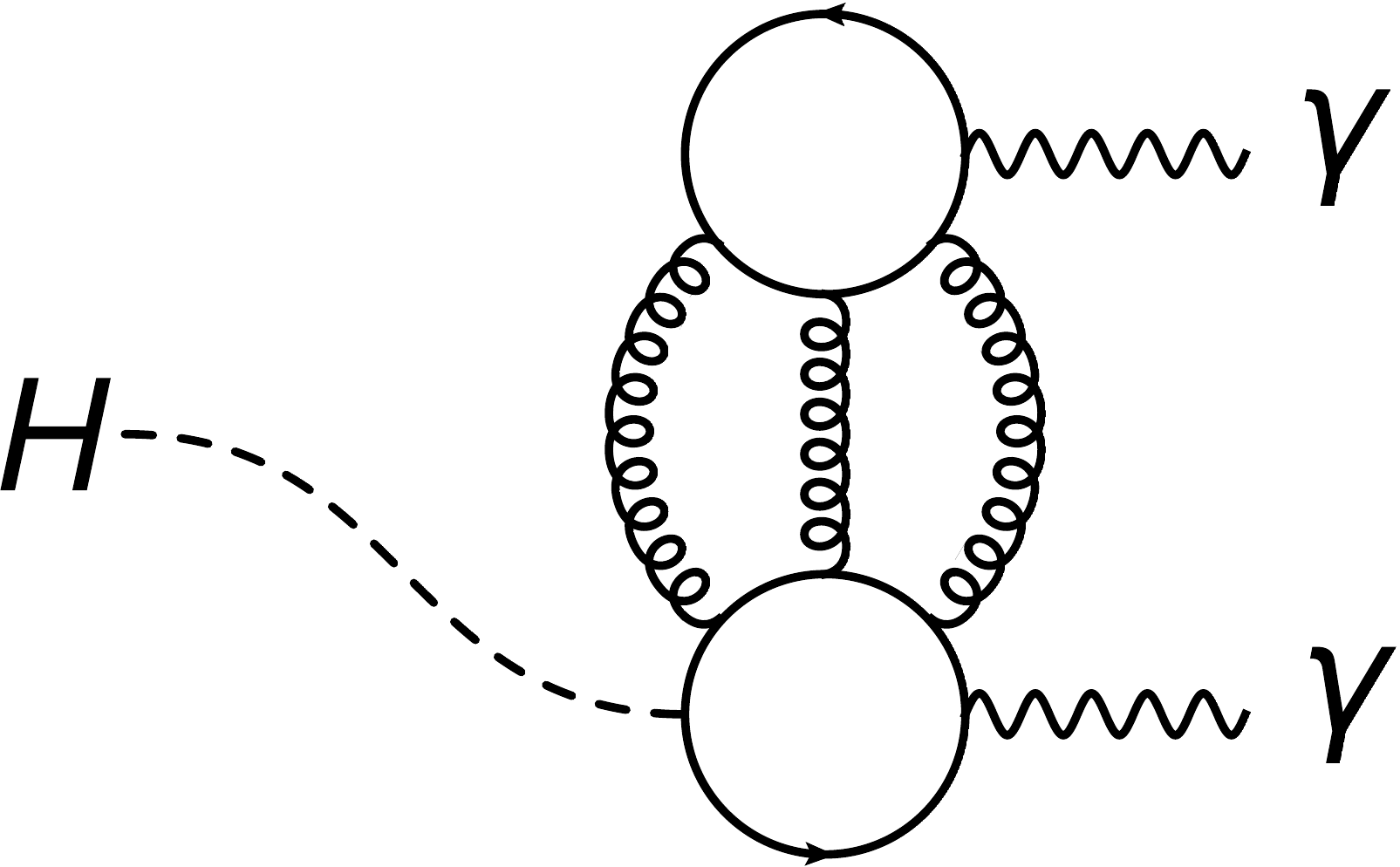}\\
\end{center}
\end{minipage}
\end{center}
\caption{Example diagrams which illustrate different kind of diagram
  classes which have been determined in the heavy top-quark mass limit
  at four-loop order. Dotted lines represent massless quarks; all other
  lines are as defined in Fig.~\ref{fig:Hgamgam3loopSing}.
\label{fig:Hgamgam4loop} }
\end{figure}

In Ref.~\cite{Chetyrkin:1997un} $C_{1\gamma}$ has been determined at
four-loop order in the effective Lagrangian approach with the help of
the knowledge of the anomalous dimensions. This result was subsequently also
obtained independently through the calculation of the four-loop order of
the decoupling function $\zeta_{g\gamma}$ in Ref.~\cite{Sturm:2014nva},
where also the corresponding five-loop contributions were determined,
with the help of the anomalous dimensions.  Some example diagrams at
four-loop order are depicted in Fig.~\ref{fig:Hgamgam4loop}.

%


\section{Conclusions\label{sec:DiscussConclude}}

The systematic investigation of four-loop tadpole integrals started about a
decade ago.
In this article we have briefly touched the techniques which have been
developed to perform the reduction to master integrals and to obtain results
for the latter. Furthermore, we have described in some detail the most
important applications. Among them are four-loop corrections to the
electroweak $\rho$ parameter, the precise determination of charm and bottom
quark masses and the decoupling constants in QCD. The latter have a close
connection to Higgs boson production and decay into gluons and photons
which has also been elaborated.


\section*{Acknowledgements}
This work is supported by the Deutsche
Forschungsgemeinschaft in the Sonderforschungsbereich Transregio~9
``Computational Particle Physics''.





\nocite{*}
\bibliographystyle{elsarticle-num}



\end{document}